\documentclass[showpacs,preprintnumbers,amsmath,10pt,a4paper,aip,jcp,reprint]{revtex4-1}

\usepackage{bm}
\usepackage{graphicx} 
\usepackage{color}
\usepackage{amsmath}
\usepackage{enumerate}

\newcommand{\EEA}{\end{eqnarray}}
\newcommand{\EEAN}{\end{eqnarray*}}

\begin{document}
 
\title{Comparison of classical and quantal calculations of helium three-body recombination}

\author{Jes\'{u}s P\'{e}rez-R\'{i}os}

\email{jperezri@purdue.edu}

\affiliation{Physics Department, 
Purdue University, West Lafayette, IN 47907,USA}

\author{Steve Ragole}

\affiliation{Joint Quantum Institute, University of Maryland, College Park, MD 20742, USA}

\author{Jia Wang}

\affiliation{Department of Physics, University of Connecticut, Storrs, CT 06269, USA }

\author{Chris H. Greene}

\affiliation{Physics Department, 
Purdue University, West Lafayette, IN 47907,USA}

\date{\today}

\begin{abstract}

A general method to study classical scattering in $n$-dimension is developed. Through classical trajectory
 calculations, the three-body recombination is computed as a function of the collision energy for helium
 atoms, as an example. Quantum calculations are also performed for the $J^{\Pi}$ = $0^{+}$ symmetry
 of the three-body recombination rate in order to compare with the classical results, yielding good 
agreement for $E\gtrsim$ 1 K. The classical threshold law is derived and numerically confirmed for
 the Newtonian three-body recombination rate. Finally, a relationship is found between the quantum and classical 
three-body hard hypersphere elastic cross sections which is analogous to the well-known shadow scattering 
in two-body collisions.

\end{abstract}

%\pacs{}

\maketitle

\section{Introduction}

True three-body collision processes (A+B+C) are vitally important in a number of physical and chemical contexts,\cite{Brown-1999, Huang-2012}, yet theoretical studies have been far less extensive than the vast literature that exists for two-body entrance channels (AB+C).  Classical trajectory methods are based on the assumption that the motion of each of the nuclei
in a chemical systems is governed to a good approximation by the laws of classical mechanics on 
the quantal potential energy surface given by the adiabatic electronic energy of the 
system. These methods have been successfully applied to calculations
of scattering observables (e.g. the differential cross section, the reaction
 probability and the integral or total cross section) in many different chemical systems, often showing good agreement with 
experimental results and with quantum 
calculations.\cite{Schneider-1995,Aoiz-1998,Aoiz-2005} 
Perhaps surprisingly, the classical trajectory methods can in some cases 
describe demanding properties of the chemical 
systems, such as the kinetic isotope effect observed in muonic 
isotopologues of the H + H$_{2}$ reaction.\cite{Jambrina-2011,Homayoon-2012}

In general, the agreement between classical trajectory methods and 
quantum mechanical calculations deteriorates at lower collision 
energies, where a low number of partial waves contribute. One might 
imagine that this issue would prevent us from studying cold and 
ultracold collisions by means of classical trajectory methods. However, in
 these situations it is informative to capitalize on the intuitive framework 
provided by the classical trajectory methods to analyze the role of the 
quantum effects and compare with Newtonian predictions.

Recently, Li and Heller\cite{Li-2012} have applied classical trajectory
 calculations to simulate buffer-gas cooling experiments with helium
 atoms and large rigid asymmetric-top molecules at a temperature as 
cold as 6.5 K. That treatment enabled the authors to calculate the lifetimes 
of long-lived orbiting resonances associated with the motion of an atom 
around a large molecule.\cite{Li-2012} They also predict that the 
production of cold molecules could be made more efficient by increasing the 
density of the buffer-gas.\cite{Li-2012} This exemplifies the potential utility of 
classical trajectory calculations for understanding the collision
 dynamics, even for cold systems.

At the same time, different mechanisms have been postulated at different times for three-body recombination.  One of these is direct three-body recombination, which is closest to the class of processes considered in the present work. A second class of mechanisms involves indirect or ``two-step’’ pathways, in which two of the particles initially collide and form a transient resonant state of vibration and/or rotation, followed by a subsequent de-excitation of the resonant state into a bound dimer in a second collision between the dimer and a third atom.  This second class has been considered extensively in the literature and there are a number of different versions of it depending on which degree of freedom is initially excited in the first resonant collision.  It is variously termed the ``Lindemann-Hinshelwood mechanism’’, or when formulated more specifically for three-atom collisions, one frequently sees reference to this as the ``RBC mechanism’’ after the authors Roberts, Bernstein, and Curtiss.\cite{Roberts-1969, Orel-1987}  

While a ``direct'' three-body recombination collision would seem to be fundamentally different from these two-step mechanisms, one study has claimed that all of the alternative three-body mechanisms are equivalent and should produce identical three-body recombination rate coefficients.\cite{Wei-1997}  The present study of helium three-body recombination can indirectly test this assertion, because the $^4$He dimer has only one bound state and no discernible resonances on the electronic ground state potential.  Thus the Lindemann-Hinshelwood or RBC mechanism predicts a vanishing three-body recombination rate, whereas both the quantal and classical calculations presented below produce direct recombination rates at low energy that exceed $10^{-30}$cm$^6$/s.  It remains unclear how this might be reconciled with the conclusions of \cite{Wei-1997}.  The long, scholarly study by Pack, Walker, and Kendrick\cite{Pack-1998} also raised doubt as to the practical applicability of this claimed equivalence of indirect versus direct mechanisms.

The continuing development of both classical\cite{Esposito-2009} and nonperturbative quantum mechanical descriptions\cite{Colavecchia-2003} of this fundamental process is also interesting in view of recent approximate theory that computes quantum mechanical ternary rates in various approximations.  One such application\cite{Forrey-2013a,Forrey-2013b} has recently been applied to the crucial astrophysical problem of hydrogen atom three-body recombination.  The hydrogen results presented appear to represent the current state of the art for three-body recombination in this system, but the main approximation utilized in that study (an energy sudden approximation) should be assessed critically. To explore other theoretical studies of recombination or its time-reverse process (collision-induced dissociation, CID) and the validity of various approximation methods, see also Refs.\cite{Ivanov-2012,Azriel-2011,Charlo-2004,Xie-2003,Bernshtein-2003,Karachevtsev-1999,Pack-1998, Colavecchia-2003,Parker-2002,Pack-1998b}

The present paper implements a general method to study classical collisions 
in $n$-dimensions. This method utilizes classical trajectory calculations to 
calculate the three-body recombination rate (TBRR) for helium atoms in a framework 
related to the pioneering formulation of F. T. Smith.\cite{Smith-1962} The collision
 treated range from ultracold energies 10$^{-4}$ K to the thermal range 10$^{3}$ K. The 
classical calculations are compared with quantum results that we have 
computed for the TBRR in the $J^{\Pi}$=$0^{+}$ symmetry only. The quantum 
calculations are based on the combination of the adiabatic hyperspherical 
representation\cite{Lin-1995,Esry-1996} and the R-matrix method to 
obtain the scattering properties.\cite{Aymar-1996} The Hamiltonian is represented in 
a discrete variable representation (DVR),\cite{Tolstikhin-1996} with some modifications 
for a better description of the nonadiabatic couplings at short-range distances.\cite{Wang-2011} 
Among the key results obtained are the classical
 threshold law for TBRR as a function of the collision 
energy and a relationship between the classical and quantum predictions 
for high collision energies. In particular, we find a result that can be interpreted 
as an $n$-dimensional generalization of the shadow scattering in two-body collisions.\cite{Child,Levine-2005}

\section{Generalized classical scattering}\label{scattering}

This section presents our method for studying classical collisions in an 
arbitrary number of dimensions $n\ge3$. This method emerges from a
 simple generalization to $n$-dimensions of the concepts necessary to
 formulate the scattering problem in classical mechanics, i.e. the impact
 parameter and the cross section. As an illustration of the method, the 
cross section for the hard-sphere model in $n$-dimensions is derived. 
This calculation generalizes the two-body hard-sphere collision 
formulas, an essential model that is useful for characterizing high energy 
collisions.

\subsection{The method}
In classical scattering, the collision between two particles can be seen in 
two equivalent ways: either a particle impinges upon a scattering center 
or two particles approach to each other with some interaction. After removing
 the ignorable center-of-mass (CM) motion and formulating the collision in the
 relative motion coordinate, the second situation is identical to the first one. This 
generalizes the notion of single particle scattering to $n$-dimensions. (For $N$
 particles colliding in ordinary three-dimensional space in the absence of an 
external field, note that the total dimensionality of the collision is equal to $n=3N-3$.)

In a three-dimensional (3D) collision, a particle with a defined momentum moves
 towards a scattering center. The cross section for a collisional process is defined
 as the area drawn in a plane perpendicular to the initial momentum, which the
 relative motion of the particles needs to cross if a collision is to take
 place.\cite{Levine-2005} The impact parameter vector, $\vec{b}$, basically 
determines the scattering, and it is defined as the component of the vector 
position that lies in the plane perpendicular to the momentum before 
the collision.\cite{Levine-2005} Both the impact parameter vector and 
the scattering cross section are readly extended to the $n$-dimensional 
scattering. However, instead of dealing with a plane we 
must work with an $n-1$-dimensional hyperplane, perpendicular to 
the direction of the initial momentum.

Thus an integral over the impact parameter will be equivalent to the 
hyperarea of the hyperplane (in 3D we use the concept of area) 
perpendicular to the initial momentum. If one attaches the opacity function 
$\wp_{process}(\vec{b},\vec{P})$ that counts the number of trajectories that 
contribute to the specific collision process of interest (e.g. chemical 
reaction, elastic scattering, etc.), then the integral will give the hyperarea 
over such hyperplane weighted by those trajectories involved in the process, 
i.e. the cross section. It can be expressed as 

\begin{equation}
\label{s-1}
\sigma_{process}(\vec{P})=\int \wp_{process}(\vec{b},\vec{P})d\Omega_{b}^{n-1}b^{n-2}db,
\end{equation} 

\noindent
where $d\Omega_{b}^{n-1}$ is the differential solid angle element of 
the hyperangles of $\vec{b}$ and $n-1$ indicates the 
dimensionality of the hyperspace where $\vec{b}$ is defined. The opacity function, 
$\wp_{\rm{process}}(\vec{b},\vec{P})$ is the probability that a trajectory with particular initial 
conditions leads to the process under study. Classical trajectory 
calculations are performed in order to determine  
the opacity function, which is the key function required in order 
to obtain the cross section.

Eq. (\ref{s-1}) yields the cross section for a particular orientation of 
the momentum, but the main quantity of interest in our present 
exploration is the cross section for a particular magnitude of the
 momentum, $P$. It is done by averaging Eq. (\ref{s-1}) over
 the hyperangles associated with the momentum, i.e.,

\begin{equation}
\label{s-2}
\sigma_{\rm{process}}(P)=\frac{\int \wp_{\rm{process}}(\vec{b},\vec{P}) d\Omega_{P}^{n}d\Omega_{b}^{n-1}b^{n-2}db}{\int d\Omega_{P}^{n}},
\end{equation} 

\noindent
where $\Omega_{P}^{n}$ represents the hyperangles of the canonical
 mass-weighted momentum vector $\vec{P}$
 in the $n$-dimensional space.

\subsection{Hard-hypersphere collision in n-dimensions}

This method is readily applied to a hard-sphere model collision in 
$n$-dimensions. We introduce 
the hard-hypersphere opacity function as 

\begin{equation}
\label{hs-1}
 \wp_{\rm{hs}}(\vec{b},\vec{P}) = \left\{ 
  \begin{array}{l l}
    1 & \quad b\le R_{0}. \\
    0 & \quad b > R_{0}. 
  \end{array} 
\right.
\end{equation}

\noindent
This expression of the opacity function is independent of the orientation 
of $\vec{b}$ and $\vec{P}$ (owing to the spherical symmetry of the model)
 and it only leads to a collision event when the magnitude of the
 impact parameter $b$ is less than a certain threshold distance $R_{0}$, the 
interaction radius of the hard-hypersphere.   

Inserting Eq.(\ref{hs-1}) into Eq. (\ref{s-2}), the cross section is expressed as 

\begin{equation}
\label{hs-2}
\sigma_{\rm{hs}}(P)=\int_{0}^{R_{0}} d\Omega_{b}^{n-1}b^{n-2}db=\Omega^{n-1}\frac{R_{0}^{n-1}}{n-1},
\end{equation} 

\noindent
where $\Omega^{n-1}$ is the solid angle subtended by a $n-1$-dimensional
 sphere (in the geometrical sense). Taking into account the general 
expression for the solid angle for a $d$-sphere 
$\Omega_{d}=\frac{2\pi^{d/2}}{\Gamma(d/2)}$,\cite{Avery} where
 $\Gamma$ is the Gamma function and $d$ represents the dimension 
of the sphere, we finally obtain

\begin{equation}
\label{hs-3}
\sigma_{\rm{hs}}(P)=\frac{2\pi^{(n-1)/2}}{\Gamma(\frac{n-1}{2})}\frac{R_{0}^{n-1}}{n-1}.
\end{equation} 
 
\noindent
Eq. (\ref{hs-3}) generalizes of the well-known expression of 
the cross section of the hard-sphere model in 3D, namely
 $\sigma=\pi R_{0}^{2}$, to $n$-dimension. To understand 
this relationship more deeply, one can visualize the circle as the 
intersection between a 3D sphere and a plane, and on the 
other hand the $n-1$-dimensional sphere is the intersection
 of an $n$-dimensional sphere and an 
$n-1$-dimensional hyperplane.

\section{Classical three-body recombination}

The $n$-dimensional classical scattering method assisted 
by classical trajectory calculations is now implemented 
to calculate the TBRR for helium atoms as a function of the
 collision energy from ultracold energies, 10$^{-4}$ K, 
to thermal energies, 1000 K. The classical trajectory calculations
 consist of a detailed implementation of the ideas pioneered by 
Felix T. Smith,\cite{Smith-1962} with Monte Carlo (MC) sampling 
of trajectories used to calculate the opacity function and the 
three-body recombination cross section (TBRCS), which leads
to the TBRR as is explained in this Section. The results we obtain
 by applying this $n$-dimensional classical scattering 
method  to three-body collisions include the classical ultracold 
threshold law for TBRR as a function of the collision energy. Another
 result that emerges is the relationship between the classical 
and quantum elastic three-body cross sections, resulting in an 
expression similar to the shadow scattering present in 
two-body collision.\cite{Child,Levine-2005}

\subsection{The equations of motion}\label{eqmot}

Next consider the classical scattering of three particles. 
Their motion of three particles 
in a potential energy landscape, 
$V(\vec{r}_{1},\vec{r}_{2},\vec{r}_{3})$ is governed by 
the Hamiltonian

\begin{equation}
\label{eq-1}
H=\frac{\vec{p}_{1}^{2}}{2m_{1}}+\frac{\vec{p}_{2}^{2}}{2m_{2}}+\frac{\vec{p}_{3}^{2}}{2m_{3}}+
V(\vec{r}_{1},\vec{r}_{2},\vec{r}_{3}),
\end{equation}

\noindent
where $\vec{p}_{i}$ and $\vec{r}_{i}$ represent the momentum 
and the vector position of the $i^{th}$ particle, respectively. In
 order to simplify the Hamiltonian we introduce the Jacobi 
coordinates (see Fig.1) defined by the relations

\begin{subequations}
\begin{eqnarray}
\vec{\rho}_{1} & = & \vec{r}_{2}-\vec{r}_{1}, \label{eq-2a}\\
\vec{\rho}_{2}&=& \vec{r}_{3}-\frac{m_{2}\vec{r}_{2}+m_{1}\vec{r}_{1}}{m_{1}+m_{2}}, \label{eq-2b}\\
\vec{\rho}_{CM}&=&\frac{m_{1}\vec{r}_{1}+m_{2}\vec{r}_{2}+m_{3}\vec{r}_{3}}{M},\label{eq-2c}
\end{eqnarray}
\end{subequations}

\noindent
where $M=m_{1}+m_{2}+m_{3}$ is the total mass of the 
system. The transformation between Cartesian coordinates 
and Jacobi ones is a contact transformation.\cite{Whittaker-1937} The 
three-body Hamiltonian can be expressed in the Jacobi 
coordinates as \cite{Karplus-1965}

\begin{equation}
\label{eq-3}
H=\frac{\vec{P}_{1}^{2}}{2m_{12}}+\frac{\vec{P}_{2}^{2}}{2m_{3,12}}+\frac{\vec{P}_{\rm{CM}}^{2}}{2M}+
V(\vec{\rho}_{1},\vec{\rho}_{2}),
\end{equation}

\noindent
where $\frac{1}{m_{12}}=\frac{1}{m_{1}}+\frac{1}{m_{2}}$; 
$\frac{1}{m_{3,12}}=\frac{1}{m_{3}}+\frac{1}{m_{1}+m_{2}}$; $V(\vec{\rho}_{1},\vec{\rho}_{2})$ 
is the potential energy expressed in terms of the Jacobi 
coordinates, which explicitly exhibits its lack of dependence 
on the center of mass (CM) coordinates, as expected, since the 
CM momentum is a constant of motion. 
$\vec{P}_{1}$,  $\vec{P}_{2}$ and $\vec{P}_{\rm{CM}}$ are the canonical 
momenta that are  conjugate to $\vec{\rho}_{1}$, $\vec{\rho}_{2}$ 
and $\vec{\rho}_{CM}$, respectively. Finally, omission of the trivial 
CM motion gives

\begin{equation}
\label{eq-4}
H=\frac{\vec{P}_{1}^{2}}{2m_{12}}+\frac{\vec{P}_{2}^{2}}{2m_{3,12}}+V(\vec{\rho}_{1},\vec{\rho}_{2}).
\end{equation}
 
  \begin{figure}[t]
\centering
	\includegraphics[width=3.0 in]{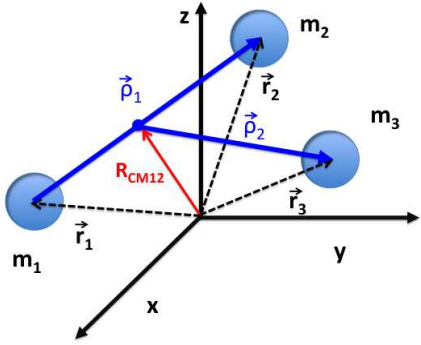}
	\caption{ (Color online) Jacobi vectors for the three-body problem}
\end{figure}

The Hamilton equations of motion for the three particles as
 a function of the Jacobi coordinates are

\begin{subequations}
\begin{eqnarray}
\frac{d\rho_{i,\alpha}}{dt}&=&\frac{\partial H}{\partial P_{i,\alpha}},\label{eq-5a} \\
\frac{dP_{i,\alpha}}{dt} &=&-\frac{\partial H}{\partial \rho_{i,\alpha}},\label{eq-5b}
\end{eqnarray}
\end{subequations}

\noindent
where $i=1,2$ and $\alpha=x,y,z$ in relation with the Cartesian
 coordinates of each Jacobi vector. The derivatives respect to 
$\rho_{i,\alpha}$ should be evaluated by taking into account 
Eqs. (\ref{eq-2a}), (\ref{eq-2b}) and  (\ref{eq-2c}), combined 
with the chain rule. The solution of Hamilton's equations
 determines the position of each particle as a
 function of time in the usual 3D space.  In the usual 
manner \cite{Smith-1962} a 6D position vector 
is constructed form the two Jacobi vectors $\vec{\rho}_{1}$ 
and $\vec{\rho}_{2}$ as

\begin{equation}
\label{eq-6}
\vec{\rho}=\begin{pmatrix} 
   \vec{\rho}_{1}\\ 
   \vec{\rho}_{2} 
\end{pmatrix}.
\end{equation}

\noindent
Hereafter, 6D vectors are represented in hyperspherical 
coordinates. The mass-scaled 6D canonical 
momentum vector can be written as  

\begin{equation}
\label{eq-7}
\vec{P}=\begin{pmatrix} 
   \sqrt{\frac{\mu}{m_{12}}}\vec{P}_{1}\\ 
   \sqrt{\frac{\mu}{m_{3,12}}}\vec{P}_{2} 
\end{pmatrix},
\end{equation}

\noindent
where $\mu=\sqrt{\frac{m_{1}m_{2}m_{3}}{M}}$. Now 
that the position vector and the momentum in a 6D 
space have been defined, the concept of impact parameter
 as the projection of the position vector in a hyperplane perpendicular to the 
initial momentum is clear. And, the three-body 
Hamiltonian is expressed as

\begin{equation}
\label{eq-8}
H=\frac{\vec{P}^{2}}{2\mu}+V(\vec{\rho}).
\end{equation}

In the present study, the TBRCS for helium atoms has 
been calculated with the three-body interaction approximated as 
the pairwise sum of helium dimer potentials, i.e.,

\begin{equation}
\label{eq-9}
V(\vec{r}_{1},\vec{r}_{2},\vec{r}_{3})=v(r_{12})+v(r_{23})+v(r_{31}),
\end{equation}

\noindent
where $r_{ij}$ are the interparticle distances. This is not an 
essential aspect of the method, however, and it 
is possible for future studies to complete these calculations
 using a full three-body potential energy surface. In particular 
these calculations adopt the helium atom-atom interaction of
 Aziz et al., designated HFD-B3-FCI1\cite{Aziz-1995}, which 
has been tested on measurements of the transport coefficients
 and the second order virial coefficient, with demonstrated accuracy
 in the predictions of this interaction potential. 

The equations of motion, Eqs. (\ref{eq-5a}) and (\ref{eq-5b}) are solved
 by means of an adaptative stepsize Runge-Kutta method, the 
Cash-Karp Runge-Kutta method.\cite{Numerical_Recipes} The time 
stepsize varies from 10$^{-7}$ ns in regions where the atomic interaction 
varies steeply, up to 10$^{-4}$ ns in regions where the interaction 
varies smoothly. The total energy is conserved during collisions to at least 
four significant digits and the total angular
 momentum, $J=|\vec{\rho}_{1}\times \vec{P}_{1}+\vec{\rho}_{2}\times \vec{P}_{2}|$ is 
conserved to at least six digits.

\subsection{Initial conditions}\label{init}

The equations of motion derived from the Hamilton's equations 
determine the time evolution of all the dynamical variables of 
the system under study, those depend on their initial
 conditions. Indeed, the selection of the initial conditions is
 a crucial step in order to characterize the scattering properties
 through the concept of impact parameter as we have seen in
 Sec. \ref{scattering}. 

The scattering of three particles can of course alternatively be viewed 
as the scattering of one particle in 6 -dimensions. 
The 6-dimensional vectors are represented
 in a hyperspherical coordinate system based on the Avery's 
definition of the hyperangles,\cite{Avery} in this representation 
all the vectors can be represented by means of their magnitude $r$ 
and five different hyperangles
 ($\alpha_{i}$, $i=1,2,3,4,5$) by

\begin{equation}
\label{init-1}
\vec{r}=\begin{pmatrix} 
   \vec{r}_{x_{1}}\\
 \vec{r}_{x_{2}}\\
\vec{r}_{x_{3}}\\
\vec{r}_{x_{4}}\\
\vec{r}_{x_{5}}\\
   \vec{r}_{x_{6}} 
\end{pmatrix}
= \begin{pmatrix} 
   r \sin{\alpha_{1}}\sin{\alpha_{2}}\sin{\alpha_{3}}\sin{\alpha_{4}}\sin{\alpha_{5}}\\
 r \cos{\alpha_{1}}\sin{\alpha_{2}}\sin{\alpha_{3}}\sin{\alpha_{4}}\sin{\alpha_{5}}\\
r \cos{\alpha_{2}}\sin{\alpha_{3}}\sin{\alpha_{4}}\sin{\alpha_{5}}\\
r \cos{\alpha_{3}}\sin{\alpha_{4}}\sin{\alpha_{5}}\\
r \cos{\alpha_{4}}\sin{\alpha_{5}}\\
  r\cos{\alpha_{5}}
\end{pmatrix}.
\end{equation}

\noindent
Here the angle ranges are $0\le\alpha_{1}\le 2\pi$, $0\le\alpha_{i}\le \pi$, $i=2,3,4,5$. In particular, we choose to 
align the 3D $z$ axis parallel to $\vec{p}_{2}$, whereby the initial momentum $\vec{P}_{0}$ is expressed as 
[see Eq. (\ref{eq-7})]

\begin{equation}
\label{init-2}
\vec{P}_{0}=\begin{pmatrix} 
   P_{0} \sin{\alpha^{P}_{1}}\sin{\alpha^{P}_{2}}\sin{\alpha^{P}_{5}}\\
 P_{0} \cos{\alpha^{P}_{1}}\sin{\alpha^{P}_{2}}\sin{\alpha^{P}_{5}}\\
P_{0} \cos{\alpha^{P}_{2}}\sin{\alpha^{P}_{5}}\\
0\\
0\\
  P_{0}\cos{\alpha^{P}_{5}}
\end{pmatrix},
\end{equation}

\noindent
 where $0\le\alpha^{P}_{1}\le 2\pi$, $0\le\alpha^{P}_{2}\le \pi$ and $0\le\alpha^{P}_{5}\le \pi$. 
The impact parameter $\vec{b}$ is defined in a 5-dimensional hyperplane orthogonal to $\vec{P}_{0}$. It is 
a 5-dimensional vector embedded in a 6-dimensional space by means of the Avery's hyperangle definition as

\begin{equation}
\label{init-3}
\vec{b}=\begin{pmatrix} 
   b \sin{\alpha^{b}_{1}} \sin{\alpha^{b}_{2}}\sin{\alpha^{b}_{3}}\sin{\alpha^{b}_{4}}\\
 b \cos{\alpha^{b}_{1}}\sin{\alpha^{b}_{2}}\sin{\alpha^{b}_{3}}\sin{\alpha^{b}_{4}}\\
b\cos{\alpha^{b}_{2}} \sin{\alpha^{b}_{3}}\sin{\alpha^{b}_{4}}\\
b\cos{\alpha^{b}_{3}}\sin{\alpha^{b}_{4}}\\
b \cos{\alpha^{b}_{4}}\\
  0 
\end{pmatrix},
\end{equation}

\noindent
where $0\le\alpha^{b}_{1}\le 2\pi$, $0\le\alpha^{b}_{i}\le \pi$, $i=2,3,4$. 

The magnitude of the initial vector position $|\vec{\rho}_{0}|=R$ is chosen to 
ensure that the interaction potential is negligible at that point and beyond. 
Then Eq. (\ref{eq-8}) yields the magnitude of the initial momentum 
$E=P_{0}^{2}/2\mu$, with $E$ the collision energy, that is equal to the kinetic 
energy initially. Our procedure randomly generates different angles $\alpha^{P}_{i}$ 
with $i=1,2,5$, and $\alpha^{b}_{j}$ with $j=1,2,3,4$, for a particular 
magnitude of the impact parameter $|\vec{b}|$. Those angles are generated by means 
of the probability density function (PDF) associated with each of them, as is show 
in Sec. \ref{opacity}. As the initial momentum $\vec{P}_{0}$ and the impact 
parameter $\vec{b}$ are orthogonal, the initial vector position can be written as 

\begin{equation}
\label{init-4}
\vec{\rho}_{0}=\vec{b}-\frac{\sqrt{R^{2}-b^{2}}}{P_{0}}\vec{P}_{0}.
\end{equation}

\noindent
Eq. (\ref{init-4}) now generates $\vec{\rho}_{0}$ from $R$, $\vec{b}$ and $\vec{P}_{0}$.

Note that the definition of the hyperradius used in this Newtonian part of the study differs from the choice of the 
hyperradius in the quantum calculations. For the classical treatment, no mass-scaling of the Jacobi coordinate vectors 
is carried out, whereas that scaling is uniformly implemented in all quantum calculations, e.g. those of this paper and 
in Refs. \cite{Esry-1999,Suno-2002,Wang-2011,Wang-2012}

\subsection{The opacity function, probability of recombination, and three-body recombination rate}\label{opacity}

The average generalized cross section for the three-body collision is given by (Eq. \ref{s-2} with $n$ = 6)

\begin{equation}
\label{op-1}
\sigma_{\rm{rec}}(P_{0})=\frac{\int \wp_{\rm{rec}}(\vec{b},\vec{P_{0}}) b^{4}db d\Omega_{b}d\Omega_{P_{0}}}{\int d\Omega_{P_{0}}}, 
\end{equation}

\noindent
This expression explicitly averages over all initial momentum directions and implies an average over
 the relative kinetic energies associated with the two initial Jacobi vectors. It takes into account the
 relationship between the initial momentum magnitude, $P_{0}$, and the 
collision energy, $E$, given in Sec. \ref{init}. $\wp_{\rm{rec}}(\vec{b},\vec{P_{0}})$ is the three-body 
recombination opacity function, which is a function of all the collisional
 parameters: $b$, $\Omega_{b}$, $P_{0}$ and $\Omega_{P_{0}}$. $\Omega_{b}$ 
and $\Omega_{P_{0}}$ represent the hyperangles associated to the vectors $\vec{b}$ and 
$\vec{P_{0}}$, respectively. $d\Omega_{b}$ and $d\Omega_{P_{0}}$ are the differential 
element of the hyperangles of each vector. In particular, the Avery hyperangle definition \cite{Avery} gives the following:

\begin{equation}
\label{op-2}
d\Omega_{b}=\sin^{3}{(\alpha^{b}_{4})}\sin^{2}{(\alpha^{b}_{3})}\sin{(\alpha^{b}_{2})}d\alpha^{b}_{4}
d\alpha^{b}_{3}d\alpha^{b}_{2}d\alpha^{b}_{1}, 
\end{equation}

\noindent
and

\begin{equation}
\label{op-3}
d\Omega_{P_{0}}=\sin^{4}{(\alpha^{P_{0}}_{5})}\sin{(\alpha^{P_{0}}_{2})}d\alpha^{P_{0}}_{5}d\alpha^{P_{0}}_{2}d\alpha^{P_{0}}_{1},
\end{equation}

\noindent
These expressions take into account the fact that we have chosen the $z$ axis parallel to the $\vec{P}_{2}$ 
direction (see Sec. \ref{init}). 

The recombination trajectories are of course those trajectories that lead to the formation of bound dimers. In classical 
trajectory methods, the states of the dimer at positive energies, which would be called shape resonances in a quantum 
calculation, appear to be truly {\it bound} dimers. Some decisions must be made as to how to take into account the 
influence of such states that we know are actually quantum mechanical shape resonances having finite 
lifetimes.\cite{Roberts-1969,Orel-1987,Esposito-2009} However, in the case of the helium dimer, we have 
checked by using the Wentzel-Kramers-Brillouin (WKB) approximation that the dimer interaction does not 
support any bound or quasi-bound state for $l\ge2$ (l is the orbital angular momentum), which confirms 
that there is no shape resonances that must be considered in the case of the helium three-body 
problem. Accordingly, the only events counted as recombination are those trajectories which lead 
to bound dimers whose energy is negative.

% In princeple we could have classical states with positive energy that do not matches neccesarily ot a shape resonance

The opacity function, $\wp_{\rm{rec}}(\vec{b},\vec{P_{0}})$, also called the reaction probability, represents the
 probability that the reactants transform into the products at each value of the initial collisional 
parameters. Usually it is given just in terms of the impact parameter $b$, but in some systems 
one is interested in the studying the effect of the reactant alignment and orientation on the
 product formation, i.e. the stereodynamics of the reaction.\cite{Levine-2005} Since the 
present study is not investigating the explicit orientation dependence of the opacity function, 
the quantity we focus on as the probability of recombination (PR) $\wp_{\rm{rec}}(b,P_{0})$ as 
the average opacity function

\begin{eqnarray}
\label{op-4}
 \wp_{\rm{rec}}(b,P_{0}) =\frac{\int  \wp_{\rm{rec}}(\vec{b},\vec{P_{0}}) d\Omega_{b}d\Omega_{P_{0}}}{\int d\Omega_{b}d\Omega_{P_{0}}}.
\end{eqnarray}

\noindent
The integral in Eq. (\ref{op-4}) is evaluated by means of the 
MC integration method,\cite{Schreiber-1966,Landau-2009} 
i.e. the initial hyperangles 
($\alpha^{b}_{4},\alpha^{b}_{3},\alpha^{b}_{2},\alpha^{b}_{1},\alpha^{P_{0}}_{5},\alpha^{P_{0}}_{1},\alpha^{P_{0}}_{2}$) 
are sampled by means of their PDF given by the Eqs. (\ref{op-2}) and (\ref{op-3}), and then the function 
$\wp_{\rm{rec}}(b,P_{0})$ is averaged over the randomly sampled hyperangles. Concretely, we run 
 $n_{\rm{tra}}(b,P_{0})$ for a given $b$ and $P_{0}$, then count the number of those trajectories that
 lead to recombination events, $n_{\rm{rec}}(b,P_{0})$, and finally the PR is calculated as  

\begin{equation}
\label{op-6}
\wp_{\rm{rec}}(b,P_{0})\approx \frac{n_{\rm{rec}}(b,P_{0})}{n_{\rm{tra}}(b,P_{0})}\pm\delta(b,P_{0}).
\end{equation} 

\noindent
Here $\delta(b,P_{0})$ is the statistical error. Since $\wp_{\rm{rec}}(b,P_{0})$ is a 
Boolean function, i.e. $<\wp^{2}_{\rm{rec}}(b,P_{0})>$ = $<\wp_{\rm{rec}}(b,P_{0})>$, 
the statistical error is given by\cite{Truhlar-1975}

\begin{equation}
\label{op-7}
\delta(b,P)=\frac{\sqrt{n_{\rm{rec}}(b,P_{0})}}{n_{\rm{tra}}(b,P_{0})}
\sqrt{\frac{n_{\rm{tra}}(b,P_{0})-n_{\rm{rec}}(b,P_{0})}{n_{\rm{tra}}(b,P_{0})}},
\end{equation}

\noindent
which assumes a one standard deviation rule (68$\%$ confidence level) for the evaluation of the 
statistical error. This error criterion is assumed for all the classical results reported in the present work.

The three-body recombination cross section is expressed in terms of the PR as

\begin{equation}
\label{op-8}
\sigma_{\rm{rec}}(P_{0})=\Omega_{b}\int_{0}^{b_{\rm{max}}(P_{0})} \wp_{\rm{rec}}(b,P_{0}) b^{4}db, 
\end{equation}

\noindent
where $b_{\rm{max}}(P_{0})$ represents the maximum impact parameter that leads to a recombination 
process for a fixed $P_{0}$, in other words, for $b>b_{max}(P_{0})$ it is found that $\wp_{\rm{rec}}(b,P_{0})=0$. 
$\Omega_{\rm{b}}=\frac{8\pi^{2}}{3}$ is the solid hyperangle associated with $\vec{b}$. The three-body 
recombination cross section is calculated by integrating Eq. (\ref{op-8}) using MC sampling of the 
magnitude of the impact parameter, $b$. Our sampling in $b$ uses the importance sampling method based on 
the weight function $e^{-b/r_{0}}$,\cite{Shui-thesis,Shui-1973} where $r_{0}$ denotes 
the range of the interaction (15 a$_{0}$ for helium dimer interaction). Then, Eq. (\ref{op-8}) is written as 

\begin{eqnarray}
\label{op-9}
\sigma_{\rm{rec}}(P_{0}) & = & \frac{8\pi^{2}}{3}\int_{0}^{b_{\rm{max}}(P_{0})}  e^{b/r_{0}}
\wp_{\rm{rec}}(b,P_{0})e^{-b/r_{0}} b^{4}db= \nonumber \\
& =& \beta\frac{{8\pi^{2}}}{3}\int_{0}^{b_{\rm{max}}(P_{0})}  e^{b/r_{0}} \wp_{\rm{rec}}(b,P_{0}) \rho(b)db
\end{eqnarray}

\noindent
where $\rho(b)=\frac{e^{-b/r_{0}} b^{4}}{\beta}$ and $\beta=\int_{0}^{b_{\rm{max}}(P_{0})}\rho(b)db$. The importance 
sampling function establishes a particular weight to any recombination trajectory as a function of $b$. In practice, 
Eq. (\ref{op-9}) is solved by sampling $b$, taking into account $\rho(b)$, and 
the initial impact parameter and momentum hyperangles
 ($\alpha^{b}_{4},\alpha^{b}_{3},\alpha^{b}_{2},\alpha^{b}_{1},\alpha^{P_{0}}_{5},\alpha^{P_{0}}_{2},\alpha^{P_{0}}_{1}$) 
by means of their PDF given by the Eqs. (\ref{op-2}) and (\ref{op-3}). After $N_{\rm{tra}}(P_{0})$ are computed and 
the number of recombination events, $N_{\rm{rec}}(P_{0})$ are counted, they are weighted by the (inverse of the) 
importance sampling function, and finally the TBRCS is calculated as

\begin{equation}
\label{op-10}
\sigma_{\rm{rec}}(P_{0})\approx \beta\frac{8\pi^{2}}{3}\frac{\sum_{i=1}^{N_{\rm{rec}}(P_{0})}e^{b_{i}/r_{0}}}{N_{\rm{tra}}(P_{0})}\pm \Delta(P_{0}),
\end{equation}

\noindent
Here $\Delta(P_{0})$ is the statistical error, given by

\begin{equation}
\label{op-11}
\Delta(P_{0})=\beta\frac{8\pi^{2}}{3}\sqrt{\frac{\frac{\sum_{i=1}^{N_{rec}(P_{0})}e^{2b_{i}/r_{0}}}{N_{\rm{tra}}(P_{0})}
-\left(\frac{\sum_{i=1}^{N_{\rm{rec}}(P_{0})}e^{b_{i}/r_{0}}}{N_{\rm{tra}}(P_{0})}\right)^{2}}{N_{\rm{tra}}(P_{0})}}.
\end{equation}

The TBRR as function of the collision energy is obtained by means of the TBRCS as

\begin{equation}
\label{op-12}
k_{3}(P_{0})=v\sigma_{\rm{rec}}(P_{0})=\frac{P_{0}}{\mu}\sigma_{\rm{rec}}(P_{0}),
\end{equation}

\noindent
where $P_{0}$ is the magnitude of the momentum in hyperspherical coordinates, which is related to 
the collision energy $E$ as shown in Sec.\ref{init}. Observe that we do compute the TBRR as 
a thermal average taking into account the Maxwell-Boltzmann weights, indeed we calculate the TBRR as a function of 
the collision energy. The thermal averaging can be readily carried out whenever it is desired to compare with the 
experiments performed in thermal equilibrium.

\subsection{The classical low energy threshold law for the three-body recombination rate}\label{threshold}

In quantum mechanics the existence of threshold laws for elastic and inelastic collisions in different 
types of collision, are very familiar, the so-called Wigner threshold laws. These threshold laws represent 
the general trend of the cross section for different processes as functions of the collision energy. In the 
classical two-body collisions between an ion and a neutral species, the Langevin cross section \cite{Levine-2005} 
can be viewed as a similar low energy threshold behavior that gives the energy dependence analytically. Next we
 show that it is possible to derive the threshold behavior of the three-neutral-body recombination rate as a function 
of the collision energy. In the low energy range where long range interactions are dominated by the van der 
Waals potential $V(R)\rightarrow C_{6}/r^{6}$. We define the maximum impact parameter $\tilde{b}_{\rm{max}}$ as the 
distance where the interaction potential is equal to the collision energy, i.e.

\begin{equation}
\label{th-1}
E=\frac{C_{6}}{\tilde{b}_{\rm{max}}^{6}}.
\end{equation}

The derivation for the classical threshold law for low energy collisions assumes a rigid-sphere 
model for the low-energy classical collisions, i.e. such that for $b<\tilde{b}_{\rm{max}}$ the
 opacity function can be taken to be 1. The three-body recombination cross section can then be expressed 
as (in virtue of Eq. (\ref{op-8}) )

\begin{equation}
\label{th-2}
\sigma_{\rm{rec}}(E)=\frac{8\pi^{2}}{3}\int_{0}^{\tilde{b}_{\rm{max}}(E)}b^{4}db\propto \tilde{b}^{5}_{\rm{max}}(E),
\end{equation}

\noindent
which has taken into account the relationship between the initial momentum and the collision energy 
(see Sec.\ref{init}). By means of Eqs. (\ref{th-1}), (\ref{th-2}) and (\ref{op-12}), and by incorporating 
the relationship between momentum and energy ($P\propto E^{1/2}$), finally we obtain 
% For Jia this paragraph is hard to follow it. We should explain it better.

\begin{equation}
\label{th-3}
k_{3}(E)\propto E^{1/2}\frac{1}{E^{5/6}}=E^{-1/3}.
\end{equation}

\subsection{Three-body elastic collisions at high collision energies}

The quantum mechanical three-body elastic cross section for distinguishable particles whose potential 
energy is short-ranged and depends on the hyperradius is given by

\begin{equation}
\label{rp-1}
\sigma^{3B} = \frac{32\pi^2}{k^5}\sum_{\lambda}N_\lambda|e^{2\imath \delta_{\lambda}}-1|^2=
\frac{32\pi^2}{k^5}4\sum_{\lambda}N_\lambda \sin^2{\delta_{\lambda}}.
\end{equation}

\noindent
In this expression $\delta_{\lambda}$ is the three-body scattering phase shift for a given value of $\lambda$, 
the eigenvalue of the squared grand angular momentum operator (see Sec. \ref{QM}), $k$ is the wave vector
 defined as $E=\hbar^{2}k^{2}/2\mu$, and $N_\lambda$ is a numerical factor that takes into account the degeneracy
 associated with the grand angular momentum

\begin{equation}
\label{rp-2}
N_{\lambda}=\frac{4+2\lambda}{4}\frac{(\lambda+3)!}{\lambda!3!}.
\end{equation}

\noindent
The summation in Eq. (\ref{rp-1}) is extended over all the values of $\lambda$ 
(partial waves). But owing to the finite range of the potential energy, $\delta_{\lambda} \rightarrow 0$ for
 high values of $\lambda$, and therefore the summation in practice has a finite number of terms. Interestingly, 
Eq. (\ref{rp-1}) exhibits a close similarity with the usual two-body classical cross 
section, where the summation is over the angular momentum $l$.\cite{Landau}

For a high energy collision a large number of partial waves must be included for the 
computation of the cross section. In general, at high energies the phase shift, $\delta_{\lambda}$, is a
rapidly varying and effectively random function of $\lambda$, whereby the three-body elastic cross 
section of Eq. (\ref{rp-1}) is given by

\begin{equation}
\label{rp-3}
\sigma^{3B} \approx
\frac{32\pi^2}{k^5}2\sum_{\lambda=0}^{\lambda_{max}}N_\lambda,
\end{equation}  

\noindent
where $\lambda_{\rm{max}}$, it is the highest value of $\lambda$ satisfying the random phase criterion. Eq. (\ref{rp-3}) 
neglects contributions to the summation for $\lambda > \lambda_{\rm{max}}$. If  $\lambda_{\rm{max}}>>1$ then

\begin{equation}
\label{rp-4}
\sum_{\lambda=0}^{\lambda_{\rm{max}}}N_{\lambda}\approx\sum_{\lambda=0}^{\lambda_{\rm{max}}} \frac{\lambda^{4}}{12}
\approx \frac{\lambda_{\rm{max}}^5}{5\times 12}.
\end{equation}

\noindent
Finally, the three-body elastic cross section is expressed as

\begin{equation}
\label{rp-5}
\sigma^{3B} \approx
\frac{16\pi^2}{3k^5}\frac{\lambda_{\rm{max}}^5}{5}.
\end{equation}

The classical three-body elastic cross section is given by Eq. (\ref{op-1}), where we have to include the right opacity function 
associated with the elastic collisions. For high energy collisions we assume a hard-sphere collision model [Eq. (\ref{hs-1})]. 
Therefore, the cross section becomes

\begin{equation}
\label{rp-6}
\sigma(P_{0})=\frac{8\pi^{2}}{3}\int_{0}^{R}b^{4}db=
\frac{8\pi^{2}R^{5}}{3\times 5}.
\end{equation}

The initial grand angular momentum is defined as 

\begin{equation}
\label{rp-7}
\vec{\Lambda}=\vec{\rho}_{0} \times \vec{P}_{0},
\end{equation}

\noindent
where the cross product is defined in a 6D space (i.e., the generalized cross product is defined in terms of the 
Levi-Civita tensor). Using the fact that $\vec{P}_{0}$ and $\vec{b}$ are orthogonal implies that

\begin{equation}
\label{rp-8}
|\vec{\Lambda}|^{2}=b^{2}P_{0}^{2}.
\end{equation}

\noindent
The eigenvalues of the grand angular momentum are $\hbar^{2}\lambda(\lambda+4)$,\cite{Avery} for high 
values of $\lambda$ which can be approximated by $\hbar^{2}\lambda^2$, thus Eq. (\ref{rp-8}) determines the 
relation between the partial wave scattering formulation and the impact parameter treatment. The result is

\begin{equation}
\label{rp-9}
\lambda_{max}=kR.
\end{equation}

\noindent
Eq. (\ref{rp-9}) establishes a relationship between the quantum three-body elastic cross section, Eq. (\ref{rp-5}), 
and the classical  Eq. (\ref{rp-6}) as

\begin{equation}
\label{rp-10}
\sigma^{3B}=2\times \sigma(P_{0}).
\end{equation}

\noindent
In Eq. (\ref{rp-10}) we notice that the high energy collision limit of the quantum three-body elastic cross section is 
two times the classical one. This same factor of two also holds for two-body collisions and it is interpreted in terms 
of shadow scattering.\cite{Child,Levine-2005} Thus, Eq. (\ref{rp-10}) implies the existence of shadow scattering in 
three-body collisions, and it counts for half of the cross section just as it does for two-body collisions.

\section{Quantum three-body recombination}\label{QM}

The present study extends the previous calculations of the TBRR for helium atoms by Suno et al.\cite{Suno-2002} 
to higher collision energies. In particular, the present quantum calculations treat only the $J^{\Pi}$ = $0^{+}$ 
symmetry and these can be compared with Newtonian recombination rate contributions for collisions with less than one unit 
of angular momentum. A different method is employed for calculating the TBRR than was by of Suno et {\it al}. The
 Schr\"{o}dinger equation is solved for three interacting helium atoms using a two steps method: first, the adiabatic
 potentials are computed by means a combination of slow variable discretization 
 (SVD) method \cite{Tolstikhin-1996} at short distances, and a traditional adiabatic hyperspherical representation
for long distances, in particular, a DVR hyperradial basis is utilized whereas the b-splines method 
is employed for the hyperangles. Second, we apply R-matrix theory\cite{Aymar-1996} for the calculation
 of the scattering properties.\cite{Wang-2011}

The three-body scattering Hamiltonian is formulated in the modified Smith-Whitten 
hyperspherical coordinates:\cite{Suno-2002} $\{R,\Omega\}\equiv \{R,\theta,\phi,\alpha,\beta,\gamma \}$. $R$ is 
the hyperradius and $\theta$ and $\phi$ are the hyperangles that describe the internal motion of the 
three-body system. In these hyperspherical coordinates the interparticle distances are expressed as\cite{Suno-2002}

\begin{subequations}
\begin{eqnarray}
r_{12} &=& 3^{-1/4}R[1+\sin{\theta}\sin{(\phi-\pi/6)}]^{1/2},\label{q-1a} \\
r_{23} &= &3^{-1/4}R[1+\sin{\theta}\sin{(\phi-5\pi/6)}]^{1/2},\label{q-1b} \\
r_{13} &=& 3^{-1/4}R[1+\sin{\theta}\sin{(\phi+\pi/2)}]^{1/2},\label{q-1c} 
\end{eqnarray}
\end{subequations}

\noindent
and the three-body Schr\"{o}dinger equation is given (in a.u.) by 

\begin{eqnarray}
\label{q-2}
&\left[ -\frac{1}{2\mu}\frac{\partial^{2}}{\partial R^{2}} +\frac{\Lambda^{2}+15/4}{2\mu R^{2}}+V(R,\theta,\phi)\right]
\psi_{\nu'}(R,\Omega) = \nonumber \\
&E\psi_{\nu'}(R,\Omega).
\end{eqnarray}

\noindent
where $\Lambda^{2}$ is the squared ''grand angular momentum operator".\cite{Suno-2002,Kendrick-1999,Lepetit-1990} 
 $\psi_{\nu'}=R^{5/2}\Psi_{\nu'}$ is a rescaled version of the usual Schr\"{o}dinger solution $\Psi_{\nu}$. The 
index $\nu'$ labels the different independent solutions that are degenerate in energy and includes the quantum numbers 
that distinguish the degenerate states.\cite{Wang-2011} The three-body interaction is taken here to be a sum of the 
three pairwise two-body helium interactions, based on the helium dimer potential of 
Aziz et {\it al.}\cite{Aziz-1995}, designated as HFD-B3-FCI1 (the same as we use for the classical calculations). Recently, Suno and Esry have performed quantum calculations 
on three-body recombination rate for helium atoms by using a potential energy surface including three-body electronic 
interactions.\cite{Suno-2008}

Eq. (\ref{q-2}) is solved in the adiabatic hyperspherical representation for a given symmetry $J^{\Pi}$. The wave 
function is represented in terms of the complete, orthonormal set of 
angular wave functions $\Phi_{\nu}(R,\Omega)$ and radial wave functions $F_{\nu \nu'}$ as

\begin{equation}
\label{q-3}
\psi_{\nu'}(R,\Omega)=\sum_{\nu=1}^{Nc}F_{\nu \nu'}(R,\Omega)\Phi_{\nu}(R,\Omega),
\end{equation}

\noindent
where $N_{c}$ indicates the number of channels in the adiabatic basis that we chose to represent the
wave function. The channel functions $\Phi_{\nu}(R,\Omega)$ are eigenfunctions of the five-dimensional partial differential 
equation

\begin{eqnarray}
\label{q-4}
&\left[ \frac{\Lambda^{2}+15/4}{2\mu R^{2}}+V(R,\theta,\phi)\right]
\Phi_{\nu}(R,\Omega) = \nonumber \\
&U_{\nu}(R)\psi_{\nu}(R,\Omega),
\end{eqnarray}

\noindent
whose solutions depend parametrically on R. Insertion of the wave function expression in terms of 
the adiabatic basis, Eq. (\ref{q-3}), into the 
Schr\"{o}dinger equation, Eq. (\ref{q-2}), yields a set of coupled ordinary differential equations in $R$

\begin{eqnarray}
\label{q-5}
\left[ -\frac{1}{2\mu}\frac{d^{2}}{d R^{2}} +U_{\nu}(R)-E\right]F_{\nu\nu'}(R) \nonumber \\
-\frac{1}{2\mu}\sum_{\gamma=1}^{N_{c}}\left[2P_{\nu\gamma}(R)\frac{d}{dR}+Q_{\nu\gamma}(R) \right]F_{\gamma\nu'}(R)=0.
\end{eqnarray}

\noindent
The matrices $P_{\nu\gamma}(R)$ and $Q_{\nu\gamma}(R)$ denote the nonadiabatic couplings between the different
adiabatic channels associated with the same symmetry. These matrices are defined as

\begin{equation}
\label{q-6}
P_{\nu\gamma}(R)=\int d\Omega \Phi_{\nu}(R,\Omega)^{*}\frac{\partial}{\partial R}\Phi_{\gamma}(R,\Omega),
\end{equation}
\noindent
and
\begin{equation}
\label{q-7}
Q_{\nu\gamma}(R)=\int d\Omega \Phi_{\nu}(R,\Omega)^{*}\frac{\partial^{2}}{\partial R^{2}}\Phi_{\gamma}(R,\Omega).
\end{equation}

\noindent
These nonadiabatic couplings are crucial in determining the inelastic transitions and the width of the resonances 
supported by the adiabatic potentials $U_{\nu}(R)$. Thus, we need to describe accurately such 
nonadiabatic couplings. In general, at short-range distances is where one would expect sharp 
nonadiabatic avoided crossings.

The solution of Eq. (\ref{q-4}) including effects of the nonadiabatic couplings 
is accomplished by means of the SVD method, which gives an efficient 
description of coupling effects.\cite{Wang-2011} Instead in the long-range
 region, the adiabatic hyperspherical coupled equations are solved directly in the 
ordinary adiabatic representation using the $P,Q$ matrices, with a DVR basis 
set.\cite{Wang-2011} This method is more practical at large $R$ where
 the nonadiabatic couplings are smooth functions of $R$. The computation 
of the nonadiabatic couplings outside of the SVD region applies the method proposed by
 Wang {\it et al.}.\cite{Wang-2012} In practice Eq. (\ref{q-4}) is solved from 1 $a_{0}$ to 35 $a_{0}$, for a set
 of 117 SVD sectors, in each of which 10 SVD points are utilized. Beyond that region, a DVR basis set is used in 
a variational R-matrix calculation with about 10$^{5}$ sectors from 35 $a_{0}$ 
to 2x10$^{4}$ $a_{0}$ for the calculation of the adiabatic potentials $U_{\nu}(R)$ 
and nonadiabatic couplings. 

The scattering S matrix is extracted from the R-matrix method in the
 long-range region. \cite{Aymar-1996} 30 adiabatic channels have been employed 
on the calculations. The solution of the adiabatic equations takes about 36 CPU hours using 
16 processors in parallel and the solution of the coupled equations is about 3 CPU hours per 
energy.

As in previous references,\cite{Esry-1999} the three-body cross section for three identical 
bosonic particles is defined using the convention of Mott and Massey,\cite{Mott} in which 
the scattering cross section $\sigma_{3}$ is the ratio of the scattered radial flux multiplied 
by $R^{5}$ divided by the incident flux in only one of the six symmetrizing permutations of the incident plane 
wave.\cite{Esry-1999,Mehta-2009} The differential cross section is then integrated over all 
final hyperangles and averaged over the initial momentum hyperangles to get an average 
generalized recombination cross section. The total three-body recombination rate is then

\begin{equation}
\label{q-8}
K_{3}=\frac{k}{\mu}\sigma_{3}^{K}=\sum_{J,\Pi}K_{3}^{J\Pi},
\end{equation}

\noindent
where $K_{3}^{J\Pi}$ is the partial recombination rate corresponding to the $J^{\Pi}$ symmetry

\begin{equation}
\label{q-9}
K_{3}^{J\Pi}=\sum_{i,f}\frac{192\left( 2J+1\right)\pi^{2}}{\mu k^{4}}|S_{i\rightarrow f}^{J\Pi}|^{2}.
\end{equation}

\noindent
Here $i$ and $f$ label the incident (three-body continuum) and outgoing (two-body 
recombination) channels, respectively, and $k=(2\mu E)^{1/2}$ are the hyperradial 
wave numbers in the incident channels. The helium dimer interaction has only one 
two-body recombination channel. Because the present study considers only the 
$J^{\Pi}=0^{+}$ symmetry, thus the in Eq. (\ref{q-8}) this summation reduces to a 
single term.

\section{Results and discussion}

The TBRR is computed as a function of the collision energy of helium atoms from 
ultracold energies up to thermal energies using both classical and quantum 
methods, and compared in the following. 

The visualization of the classical trajectories helps to understand the physical 
mechanism controlling the recombination process being investigated. Fig. 2 
shows two recombination trajectories at 10$^{-4}$ K for different impact 
parameters: null impact parameter (upper panel), and $b$ = 100 a$_{0}$
 (lower panel). Observe that the collisions for different impact parameters 
lead to recombination trajectories with different final states of rotation and 
vibration of the {\it bound} dimers. Thus, the energy of the dimers associated with 
those trajectories will be different. We note that the trajectories with null 
impact parameter start to be deflected at distances where one of the 
interatomic distance has a small value. On the contrary, the trajectories 
at high impact parameter show a deflection before any interatomic distance
 start to be small. Fig. 2 shows that the three-body recombination events 
need at least two atom-atom collisions in order to form a {\it bound} dimer. It
is instructive to introduce the three-body collision time, $\tau_{3}$, as the time 
from when the atoms first start to experience their interaction (i.e. as reflected by 
a change in the initial slope of the trajectory ) until the time when two atoms 
form a {\it bound} dimer. Analysis of Fig. 2 shows that the collision time is 
$\tau_{3} \sim 10$ ns at 10$^{-4}$ K. At 1000 K the three-body collision time 
for a characteristic trajectory is $\tau_{3} \sim 10^{-3}$ ns.

  \begin{figure}[t]
\centering
	\includegraphics[width=2.5 in]{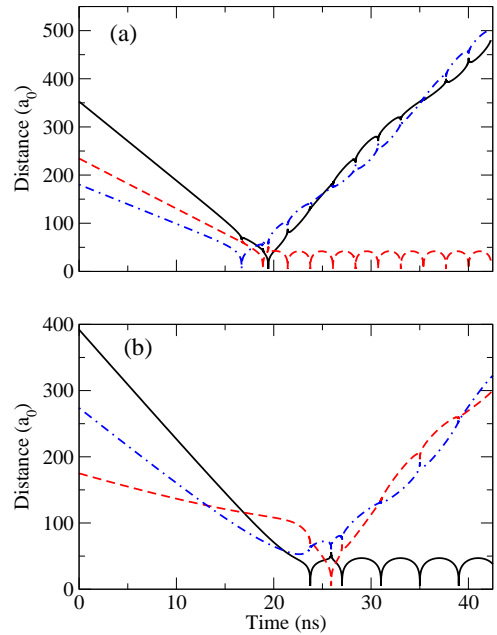}
	\caption{(Color online) Examples of recombination trajectories for
 E = 10$^{-4}$ K. Panel (a): classical trajectories for $b$=0 a$_{0}$. 
Panel (b), classical trajectories for $b$=100 a$_{0}$. The solid line represents
 $r_{12}$, the dashed line represents $r_{23}$ and the dot-dashed line represents $r_{31}$}
\end{figure}

The probability of recombination (PR) [$\wp_{\rm{rec}}(b,P)$] for helium atoms as a 
function of the impact parameter, at collisional energies: 10$^{-4}$ K, 10$^{-2}$ K, 
1 K and 25 K, is presented in Fig. 3. The PR has been calculated following the method 
described in Sec. \ref{opacity}. These results use an equally spaced grid in $b$, and for 
each of those $b$-values we run 10$^{3}$ trajectories for the calculation of 
the PR [Eq. (\ref{op-6})] and its statistical error [Eq. (\ref{op-7})], i.e. 
$n_{\rm{tra}}(b,P_{0})=10^{3}$. From this figure the maximum impact parameter producing 
a recombination for a given energy, $b_{\rm{max}}$, is extracted, and this equals: 
235 a$_{0}$, 120 a$_{0}$, 58 a$_{0}$ and 40 a$_{0}$, for 10$^{-4}$ K, 
10$^{-2}$ K, 1 K and 25 K, respectively. We note a general trend for $E<1$ K: a reduction 
of $b_{\rm{max}}$ by a factor of two while the change in energy is two orders of magnitude. 
However, this trend for $b_{\rm{max}}$ changes for energies between 1 K and 25 K, where it 
exhibits a steep trend as a function of the collision energy. This suggests that a different 
dynamical regime starts to play a role for $E\gtrsim1$ K.

  \begin{figure}[t]
\centering
	\includegraphics[width=2.5 in]{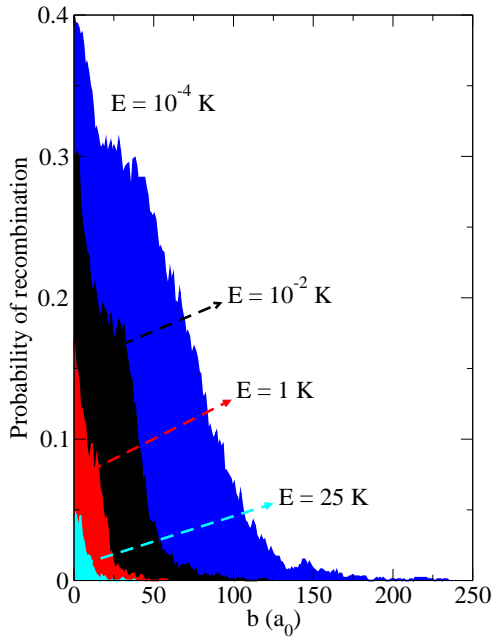}
	\caption{(Color online) Probability of recombination for helium atoms as a function of the 
impact parameter, at different collision energies: blue line, E = 10$^{-4}$ K; black line, 
E = 10$^{-2}$ K; red line, E = 1 K; cyan line, E = 25 K. }
\end{figure}

The TBRR for helium atoms as a function of the collision energy is presented in 
Fig. 4. The classical calculations have been performed following the methodology
developed in Sec. \ref{opacity}. Batches of 10$^{5}$ trajectories have been computed for
 each collision energy, except for $E\ge 100$ K, for those energies 10$^{6}$
trajectories are computed in order to have a reliable statistics in the 
three-body recombination events. The partial three-body rate for total angular
 momentum $J=0$ is also displayed. It is calculated by counting only those recombination 
trajectories having total angular momentum between $J=0$ and $J=\hbar$ and 
then applying Eqs. (\ref{op-10}) and (\ref{op-11}). The quantum mechanical calculation 
for $J^{\Pi}$ = $0^{+}$ has been carried out by means of the adiabatic hyperspherical 
representation outlined in Sec. \ref{QM}. The quantum calculation has been divided by the 
symmetrization factor $3!$ in order to compare with the classical partial three-body 
rate; the $3!$ is a factor associated with the quantal indistinguishability of the colliding
 particles, and it is divided out to compare with the classical theory which of course treats the three particles as inherently distinguishable. This figure also shows the results of the three-body recombination 
rate coefficient for helium atoms deduced in a theoretical model based on the a 
simulation of the experimental conditions.\cite{Bruch-2002} This model has some
 inputs that correspond with experimental values, whereas it has free 
parameters that are model dependent. The present calculations fix the scattering length 
of the helium dimer interaction, $a=172$ a$_{0}$\cite{Suno-2002} 
and $\phi_{b}$=47 (see Ref. \cite{Bruch-2002} for details of their model).

  \begin{figure}[t]
\centering
	\includegraphics[width=2.5 in]{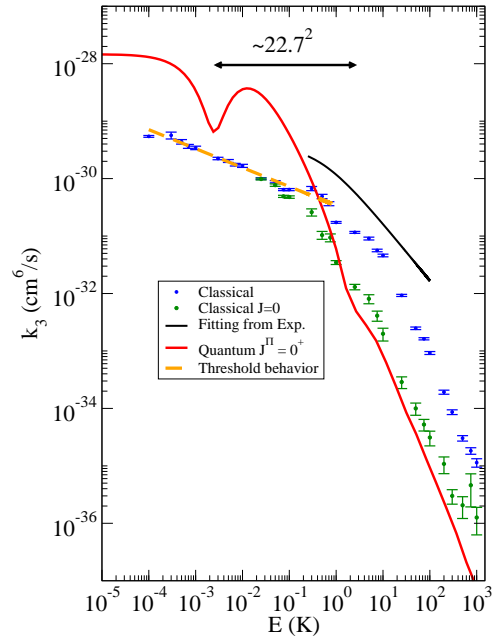}
	\caption{(Color online) The three-body recombination rate of helium atoms 
(in cm$^{6}/$s) computed using different methods is shown as a function of the collision 
energy $E$ (in K). Solid red line: quantum calculation for a fixed total angular 
momentum and parity $J^{\Pi}$=0$^{+}$. This calculations show convergence to within
better than about 15$\%$ for $E$=1000 K. Blue points: classical trajectory 
calculations. Dark green points: classical trajectory calculations for a fixed 
total angular momentum $J$=0. Solid black line: theoretical model
 of Ref. \cite{Bruch-2002}. }
\end{figure}

Compare first the classical partial three-body rate for $J$ = 0 
with the quantum calculation for the same total angular momentum. 
These calculations show good agreement for collision energies 
$E\gtrsim 1K$. The van der Waals energy 
$E_{\rm{vdW}}=\left(2\hbar^{6}/\mu^{3}C_{6}\right)^{1/2}$ is defined as a 
threshold energy between the ultracold physics $E$ < $E_{vdW}$ and 
the thermal physics $E > E_{\rm{vdW}}$.\cite{Jones-2006} For the helium 
dimer potential used in our calculations the van der Waals 
energy is $E_{\rm{vdW}}$ = 1.68 K, which marks the lowest energy where 
quantum and classical calculations begin to show the same trend, even 
though they are not yet in quantitative agreement at that energy.  

Naturally the classical calculation exhibits no sign of the quantal interference minimum that is prominent around 2-4 mK.  This quantal minimum is a universal feature associated with Efimov physics, as has been discussed previously.\cite{Suno-2002,Shepard-2007,Rittenhouse-2011JPB}  Esry and coworkers have stressed that the log-periodic behavior of Efimov physics, which originates in the attractive $1/r^2$ effective potential, should be observable in energy-dependences in addition to the more usual abscissa in ultracold experiments which is the scattering length.  The small inflection in the quantal calculation around 20 K in Fig.4 is presumably the expected second interference feature, since it is seen close to where it is expected to appear, namely at an energy $22.7^2=515$ times the energy of the first interference minimum.\cite{DIncao-2013AdvAtMolecPhys}

The classical calculations for three-body recombination rate (blue circles) show 
fair qualitative agreement with the three-body recombination rate coefficient
 (solid black line) obtained in the model of Bruch {\it et al.}.\cite{Bruch-2002} It is 
particulary interesting that the predictions of the model as well as the classical 
calculations show a change in the trend of the curves at around the same energy 
collision energy. We emphasize that the classical calculations are energy 
dependent, i.e. no Maxwell-Boltzmann thermal averaging has been performed, whereas 
the model of  Bruch {\it et al.} is based on a thermally averaged recombination rate 
coefficient.

  \begin{figure}[t]
\centering
	\includegraphics[width=2.5 in]{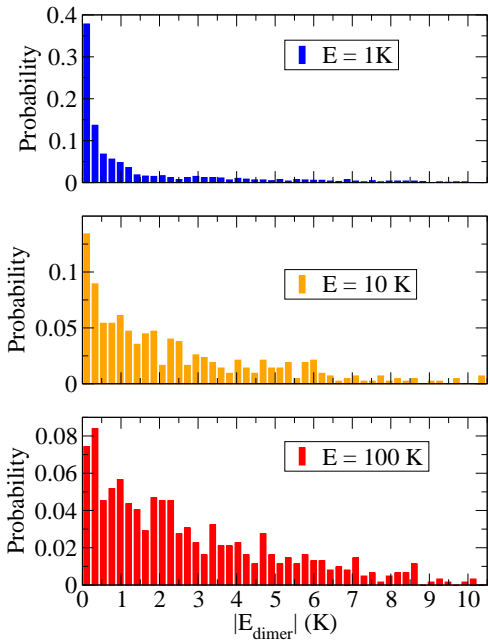}
	\caption{(Color online) Energy distribution of the helium dimers formed through three-body recombination trajectories at 
different collision energies: $E$ = 1 K, blue bars; $E$ = 10 K, orange bars; $E$ = 100 K, red bars.  }
\end{figure}

Next, consider the behavior of the classical TBRR as a function of the 
collision energy for low energies $E\lesssim 1$ K. The trend 
of the classical calculations is well described by the orange 
dashed line. This line is a fit of the classical calculations to the classical
 ultracold threshold law for TBRR [Eq. (\ref{th-3})] derived above in the 
$n$-dimensional scattering method  of Sec. \ref{threshold}. As an 
independent test not related to our present calculation for helium, we 
have performed calculations for the TBRR as a function of the collision 
energy using a Lennard-Jones interaction between the atoms, and this 
test again agrees with the classical low energy threshold law.  This
 numerical demonstration of the classical low energy threshold law confirms 
that the long-range interaction controls the low energy limit of classical 
three-body recombination.

The classical three-body rate behaves differently at $E\gtrsim 10$ K than 
it does at low collision energies. For collision energies higher than the the well
 depth, the classical trajectories are mainly deflected by the short-range part 
of the interaction, i.e. the hard inner wall of the two-body interaction starts to
 play a role. The helium dimer potential employed in our calculations has a well
 depth $\approx$11 K, which could explain the change in the trend of the classical 
calculation observed at that energy in Fig. 4.

In this vein, consider the energy distribution of the dimers formed in three-body
 recombination of helium atoms. This is presented in Fig. 5 as histogram of the 
dimer energies resulting from the three-body collisions. The distribution of dimer 
energies is observed to vary as one increases the collision energy. The energy 
distribution goes from a sharp distribution ($E$ = 1 K) to a flatter one 
($E$ = 100 K). It is related to the role of the two-body 
interaction: for long-range dominated collisions the recombination processes 
start to occur at long distances between the atoms, leading mainly to the formation
 of shallow dimers. However, as one increases the collision energy, the recombination 
process will occur in inner regions of the atom-atom interaction, leading to an 
increasing of the probability to form dimers with higher binding energy.

  \begin{figure}[t]
\centering
	\includegraphics[width=2.5 in]{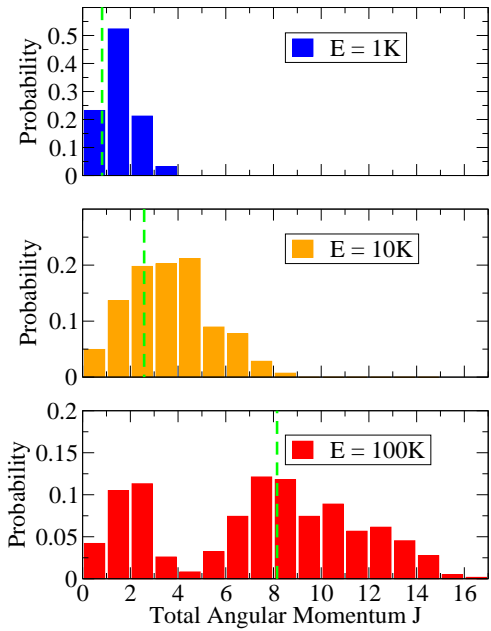}
	\caption{(Color online) Total angular momentum $J$ distribution associated with
 the recombination trajectories at three collision energies: $E$ = 1 K, blue bars; 
$E$ = 10 K, orange bars; $E$ = 100 K, red bars.  }
\end{figure}

The distribution of total angular momentum $J$ of the classical recombination 
trajectories for different collision energies is plotted as a histogram in Fig. 6. 
 This figure also plots the total angular momentum 
associated with the van der Waals length 
$R_{\rm{vdw}}=1/2\left(2\mu C_{6}/\hbar^{2} \right)^{1/4}$, which suggests a maximum 
value of the total angular momentum of the system contributing to recombination 
collisions if they are entirely dominated by the long-range interaction. It may be 
estimated as $J_{\rm{vdW}}\sim kR_{\rm{vdW}}$, where $R_{\rm{vdW}}$ is the van der Waals length 
and $k$ is the hyperspherical wave number. For the helium dimer interaction employed 
in this work it is $R_{\rm{vdW}}\approx$ 5 a$_{0}$. The results show good agreement 
between the most probable total angular momentum of the recombination trajectories 
and $J_{\rm{vdW}}$, except for the distribution at $E$ = 100 K, which shows a bimodal 
distribution. This bimodal distribution is presumably correlated with the change in 
the dimer energy distribution. For higher collision energies the probability to form 
dimers with high binding energy (small size) increases (see Fig. 5), so the contribution
 of small dimers will show up as a peak in the low total angular momentum. Indeed, this 
was noted in our discussion of Fig. 6. Finally, the change of trend in the classical 
three-body rate at $E\gtrsim 10$ K is evidently due to the increasing importance of 
the short-range interaction beyond that energy range. Moreover, the hard inner wall is
 responsible for the energy distribution of the dimers at high collision energies, 
and presumably for the bimodal distribution that deviates from the van der Waals 
estimation $J_{\rm{vdW}}$.

\section{Summary}

The present investigation has developed a general method to study classical scattering
 in $n$-dimensions. Application of this method to three-body collisions has shown a 
number of useful and interesting results: the classical ultracold threshold law for 
TBRR as a function of the collision energy and a high energy relationship between the classical three-body 
elastic cross section an the quantum one. The former has been shown numerically, the latter show a similar 
relationship that the two-body collision, leading to generalize the shadow scattering phenomena to 
three-body collisions.

The TBRR as a function of the collision energy for helium atoms has been calculated 
by the $n$-dimensional scattering method using classical trajectories. The general trends
 of the classical TBRR have been established the different influences of the long- versus 
short-range portions of the two-body interaction, as the collision energy changes. The 
classical results have been tested against our quantum mechanical TBRR, and the classical 
calculations accurately reproduce the trend of the quantum mechanical calculations for
 $E\gtrsim 1$ K. Nevertheless, quantitative agreement of the rates is not achieved, as a 
residual discrepancy between the high energy limits of classical and quantum 
TBRR is seen to be approximately.

Although the classical results have not quantitatively replicated the quantum mechanical
 results, especially at very low energies, the Newtonian treatment is a useful tool that can help in understanding 
three-body collisions. For instance, classical trajectory calculations provide a first 
estimate of the number of partial waves needed for a quantum mechanical calculation in
the thermal regime. Also, use of classical trajectory methods to determine the energy 
distribution of the dimers formed allows an estimation of which dimer levels are the 
most important for quantum calculations. Finally, the interpretation of the TBRR 
through classical trajectory calculations gives important clues that can give insight 
into mechanisms relevant for the quantum mechanical calculations.

\section{Acknowledgements}
We are indebted to Jose D'Incao for his assistance and advice regarding the calculations.  This work is supported in part by the U.S. Department of Energy, Office of Science.

%\bibliography{He-recomb}

\begin{thebibliography}{56}
\expandafter\ifx\csname natexlab\endcsname\relax\def\natexlab#1{#1}\fi
\expandafter\ifx\csname bibnamefont\endcsname\relax
  \def\bibnamefont#1{#1}\fi
\expandafter\ifx\csname bibfnamefont\endcsname\relax
  \def\bibfnamefont#1{#1}\fi
\expandafter\ifx\csname citenamefont\endcsname\relax
  \def\citenamefont#1{#1}\fi
\expandafter\ifx\csname url\endcsname\relax
  \def\url#1{\texttt{#1}}\fi
\expandafter\ifx\csname urlprefix\endcsname\relax\def\urlprefix{URL }\fi
\providecommand{\bibinfo}[2]{#2}
\providecommand{\eprint}[2][]{\url{#2}}

\bibitem[{\citenamefont{Brown et~al.}(1994)\citenamefont{Brown, Taulkdar, and
  Ravishankara}}]{Brown-1999}
\bibinfo{author}{\bibfnamefont{S.~S.} \bibnamefont{Brown}},
  \bibinfo{author}{\bibfnamefont{R.~K.} \bibnamefont{Taulkdar}},
  \bibnamefont{and} \bibinfo{author}{\bibfnamefont{A.~R.}
  \bibnamefont{Ravishankara}}, \bibinfo{journal}{Chem. Phys. Lett.}
  \textbf{\bibinfo{volume}{299}}, \bibinfo{pages}{277} (\bibinfo{year}{1994}).

\bibitem[{\citenamefont{Huang et~al.}(2012)\citenamefont{Huang, Eskola, and
  Taatjes}}]{Huang-2012}
\bibinfo{author}{\bibfnamefont{H.}~\bibnamefont{Huang}},
  \bibinfo{author}{\bibfnamefont{A.~J.} \bibnamefont{Eskola}},
  \bibnamefont{and} \bibinfo{author}{\bibfnamefont{C.~A.}
  \bibnamefont{Taatjes}}, \bibinfo{journal}{J. Phys. Chem. Lett.}
  \textbf{\bibinfo{volume}{3}}, \bibinfo{pages}{3399} (\bibinfo{year}{2012}).

\bibitem[{\citenamefont{Schneider et~al.}(1995)\citenamefont{Schneider,
  Seekamp-Rahn, Borkowski, Wrede, Welge, Aoiz, Ba\~nares, D\'{}Mello, Herrero,
  S\'{a}ez~R\'{a}banos et~al.}}]{Schneider-1995}
\bibinfo{author}{\bibfnamefont{L.}~\bibnamefont{Schneider}},
  \bibinfo{author}{\bibfnamefont{D.}~\bibnamefont{Seekamp-Rahn}},
  \bibinfo{author}{\bibfnamefont{J.}~\bibnamefont{Borkowski}},
  \bibinfo{author}{\bibfnamefont{E.}~\bibnamefont{Wrede}},
  \bibinfo{author}{\bibfnamefont{K.~H.} \bibnamefont{Welge}},
  \bibinfo{author}{\bibfnamefont{F.~J.} \bibnamefont{Aoiz}},
  \bibinfo{author}{\bibfnamefont{L.}~\bibnamefont{Ba\~nares}},
  \bibinfo{author}{\bibfnamefont{M.~J.} \bibnamefont{D\'{}Mello}},
  \bibinfo{author}{\bibfnamefont{V.~J.} \bibnamefont{Herrero}},
  \bibinfo{author}{\bibfnamefont{V.}~\bibnamefont{S\'{a}ez~R\'{a}banos}},
  \bibnamefont{et~al.}, \bibinfo{journal}{Science}
  \textbf{\bibinfo{volume}{269}}, \bibinfo{pages}{207} (\bibinfo{year}{1995}).

\bibitem[{\citenamefont{Aoiz et~al.}(1998)\citenamefont{Aoiz, Ba\~nares, and
  Herrero}}]{Aoiz-1998}
\bibinfo{author}{\bibfnamefont{F.~J.} \bibnamefont{Aoiz}},
  \bibinfo{author}{\bibfnamefont{L.}~\bibnamefont{Ba\~nares}},
  \bibnamefont{and} \bibinfo{author}{\bibfnamefont{V.~J.}
  \bibnamefont{Herrero}}, \bibinfo{journal}{J. Chem. Soc. Faraday Trans.}
  \textbf{\bibinfo{volume}{94}}, \bibinfo{pages}{2483} (\bibinfo{year}{1998}).

\bibitem[{\citenamefont{Aoiz et~al.}(2005)\citenamefont{Aoiz,
  S\'{a}ez-R\'{a}banos, Mart\'{i}nez-Haya, and
  Gonz\'{a}lez-Lezana}}]{Aoiz-2005}
\bibinfo{author}{\bibfnamefont{F.~J.} \bibnamefont{Aoiz}},
  \bibinfo{author}{\bibfnamefont{V.}~\bibnamefont{S\'{a}ez-R\'{a}banos}},
  \bibinfo{author}{\bibfnamefont{B.}~\bibnamefont{Mart\'{i}nez-Haya}},
  \bibnamefont{and}
  \bibinfo{author}{\bibfnamefont{T.}~\bibnamefont{Gonz\'{a}lez-Lezana}},
  \bibinfo{journal}{J. Chem. Phys.} \textbf{\bibinfo{volume}{123}},
  \bibinfo{pages}{094101} (\bibinfo{year}{2005}).

\bibitem[{\citenamefont{Jambrina et~al.}(2011)\citenamefont{Jambrina,
  Garc\'{i}a, Herrero, S\'{a}ez-R\'{a}banos, and Aoiz}}]{Jambrina-2011}
\bibinfo{author}{\bibfnamefont{P.~G.} \bibnamefont{Jambrina}},
  \bibinfo{author}{\bibfnamefont{E.}~\bibnamefont{Garc\'{i}a}},
  \bibinfo{author}{\bibfnamefont{V.~J.} \bibnamefont{Herrero}},
  \bibinfo{author}{\bibfnamefont{V.}~\bibnamefont{S\'{a}ez-R\'{a}banos}},
  \bibnamefont{and} \bibinfo{author}{\bibfnamefont{F.~J.} \bibnamefont{Aoiz}},
  \bibinfo{journal}{J. Chem. Phys.} \textbf{\bibinfo{volume}{135}},
  \bibinfo{pages}{034310} (\bibinfo{year}{2011}).

\bibitem[{\citenamefont{Homayoon et~al.}(2012)\citenamefont{Homayoon, Jambrina,
  Aoiz, and Bowman}}]{Homayoon-2012}
\bibinfo{author}{\bibfnamefont{Z.}~\bibnamefont{Homayoon}},
  \bibinfo{author}{\bibfnamefont{P.~G.} \bibnamefont{Jambrina}},
  \bibinfo{author}{\bibfnamefont{F.~J.} \bibnamefont{Aoiz}}, \bibnamefont{and}
  \bibinfo{author}{\bibfnamefont{J.~M.} \bibnamefont{Bowman}},
  \bibinfo{journal}{J. Chem. Phys.} \textbf{\bibinfo{volume}{137}},
  \bibinfo{pages}{021102} (\bibinfo{year}{2012}).

\bibitem[{\citenamefont{Li and Heller}(2012)}]{Li-2012}
\bibinfo{author}{\bibfnamefont{Z.}~\bibnamefont{Li}} \bibnamefont{and}
  \bibinfo{author}{\bibfnamefont{E.~J.} \bibnamefont{Heller}},
  \bibinfo{journal}{J. Chem. Phys.} \textbf{\bibinfo{volume}{136}},
  \bibinfo{pages}{054306} (\bibinfo{year}{2012}).

\bibitem[{\citenamefont{Roberts et~al.}(1969)\citenamefont{Roberts, Bernstein,
  and Curtiss}}]{Roberts-1969}
\bibinfo{author}{\bibfnamefont{R.~E.} \bibnamefont{Roberts}},
  \bibinfo{author}{\bibfnamefont{R.~B.} \bibnamefont{Bernstein}},
  \bibnamefont{and} \bibinfo{author}{\bibfnamefont{C.~F.}
  \bibnamefont{Curtiss}}, \bibinfo{journal}{J. Chem. Phys.}
  \textbf{\bibinfo{volume}{50}}, \bibinfo{pages}{5163} (\bibinfo{year}{1969}).

\bibitem[{\citenamefont{Orel}(1987)}]{Orel-1987}
\bibinfo{author}{\bibfnamefont{A.~E.} \bibnamefont{Orel}}, \bibinfo{journal}{J.
  Chem. Phys.} \textbf{\bibinfo{volume}{87}}, \bibinfo{pages}{314}
  (\bibinfo{year}{1987}).

\bibitem[{\citenamefont{Wei et~al.}(1997)\citenamefont{Wei, Alavi, and
  Snider}}]{Wei-1997}
\bibinfo{author}{\bibfnamefont{G.~W.} \bibnamefont{Wei}},
  \bibinfo{author}{\bibfnamefont{S.}~\bibnamefont{Alavi}}, \bibnamefont{and}
  \bibinfo{author}{\bibfnamefont{R.~F.} \bibnamefont{Snider}},
  \bibinfo{journal}{J. Chem. Phys.} \textbf{\bibinfo{volume}{106}},
  \bibinfo{pages}{1463} (\bibinfo{year}{1997}).

\bibitem[{\citenamefont{Pack et~al.}(1998{\natexlab{a}})\citenamefont{Pack,
  Walker, and Kendrick}}]{Pack-1998}
\bibinfo{author}{\bibfnamefont{R.~T.} \bibnamefont{Pack}},
  \bibinfo{author}{\bibfnamefont{R.~B.} \bibnamefont{Walker}},
  \bibnamefont{and} \bibinfo{author}{\bibfnamefont{B.~K.}
  \bibnamefont{Kendrick}}, \bibinfo{journal}{J. Chem. Phys.}
  \textbf{\bibinfo{volume}{109}}, \bibinfo{pages}{6701}
  (\bibinfo{year}{1998}{\natexlab{a}}).

\bibitem[{\citenamefont{Esposito and Capitelli}(2009)}]{Esposito-2009}
\bibinfo{author}{\bibfnamefont{F.}~\bibnamefont{Esposito}} \bibnamefont{and}
  \bibinfo{author}{\bibfnamefont{M.}~\bibnamefont{Capitelli}},
  \bibinfo{journal}{J. Phys. Chem. A} \textbf{\bibinfo{volume}{113}},
  \bibinfo{pages}{15307} (\bibinfo{year}{2009}).

\bibitem[{\citenamefont{Colavecchia et~al.}(2003)\citenamefont{Colavecchia,
  Mrugala, Parker, and Pack}}]{Colavecchia-2003}
\bibinfo{author}{\bibfnamefont{F.~D.} \bibnamefont{Colavecchia}},
  \bibinfo{author}{\bibfnamefont{F.}~\bibnamefont{Mrugala}},
  \bibinfo{author}{\bibfnamefont{G.~A.} \bibnamefont{Parker}},
  \bibnamefont{and} \bibinfo{author}{\bibfnamefont{R.~T.} \bibnamefont{Pack}},
  \bibinfo{journal}{J. Chem. Phys.} \textbf{\bibinfo{volume}{118}},
  \bibinfo{pages}{10387} (\bibinfo{year}{2003}).

\bibitem[{\citenamefont{Forrey}(2013)}]{Forrey-2013a}
\bibinfo{author}{\bibfnamefont{R.~C.} \bibnamefont{Forrey}},
  \bibinfo{journal}{Astrophys. J. Lett.} \textbf{\bibinfo{volume}{773}},
  \bibinfo{pages}{L25} (\bibinfo{year}{2013}).

\bibitem[{\citenamefont{Forrey}()}]{Forrey-2013b}
\bibinfo{author}{\bibfnamefont{R.~C.} \bibnamefont{Forrey}},
  \bibinfo{howpublished}{arXiv:1307.0549}.

\bibitem[{\citenamefont{Ivanov and Babikov}(2012)}]{Ivanov-2012}
\bibinfo{author}{\bibfnamefont{M.~V.} \bibnamefont{Ivanov}} \bibnamefont{and}
  \bibinfo{author}{\bibfnamefont{D.}~\bibnamefont{Babikov}},
  \bibinfo{journal}{J. Chem. Phys.} \textbf{\bibinfo{volume}{136}},
  \bibinfo{pages}{184304} (\bibinfo{year}{2012}).

\bibitem[{\citenamefont{Azriel et~al.}(2011)\citenamefont{Azriel, Kolesnikova,
  Yu~Rusin, and Sevryuk}}]{Azriel-2011}
\bibinfo{author}{\bibfnamefont{V.~M.} \bibnamefont{Azriel}},
  \bibinfo{author}{\bibfnamefont{E.~V.} \bibnamefont{Kolesnikova}},
  \bibinfo{author}{\bibfnamefont{L.}~\bibnamefont{Yu~Rusin}}, \bibnamefont{and}
  \bibinfo{author}{\bibfnamefont{M.~B.} \bibnamefont{Sevryuk}},
  \bibinfo{journal}{J. Phys. Chem. A} \textbf{\bibinfo{volume}{115}},
  \bibinfo{pages}{7055} (\bibinfo{year}{2011}).

\bibitem[{\citenamefont{Charlo and Clary}(2004)}]{Charlo-2004}
\bibinfo{author}{\bibfnamefont{D.}~\bibnamefont{Charlo}} \bibnamefont{and}
  \bibinfo{author}{\bibfnamefont{D.}~\bibnamefont{Clary}}, \bibinfo{journal}{J.
  Chem. Phys.} \textbf{\bibinfo{volume}{120}}, \bibinfo{pages}{2700}
  (\bibinfo{year}{2004}).

\bibitem[{\citenamefont{Xie et~al.}(2003)\citenamefont{Xie, Poirier, and
  Gellene}}]{Xie-2003}
\bibinfo{author}{\bibfnamefont{J.}~\bibnamefont{Xie}},
  \bibinfo{author}{\bibfnamefont{B.}~\bibnamefont{Poirier}}, \bibnamefont{and}
  \bibinfo{author}{\bibfnamefont{G.~I.} \bibnamefont{Gellene}},
  \bibinfo{journal}{J. Chem. Phys.} \textbf{\bibinfo{volume}{119}},
  \bibinfo{pages}{10678} (\bibinfo{year}{2003}).

\bibitem[{\citenamefont{Bernshtein and Oref}(2003)}]{Bernshtein-2003}
\bibinfo{author}{\bibfnamefont{V.}~\bibnamefont{Bernshtein}} \bibnamefont{and}
  \bibinfo{author}{\bibfnamefont{I.}~\bibnamefont{Oref}}, \bibinfo{journal}{J.
  Chem. Phys.} \textbf{\bibinfo{volume}{118}}, \bibinfo{pages}{10611}
  (\bibinfo{year}{2003}).

\bibitem[{\citenamefont{Karachevtsev and Vinogradov}(1999)}]{Karachevtsev-1999}
\bibinfo{author}{\bibfnamefont{G.~V.} \bibnamefont{Karachevtsev}}
  \bibnamefont{and} \bibinfo{author}{\bibfnamefont{P.~S.}
  \bibnamefont{Vinogradov}}, \bibinfo{journal}{Russian Rev. Chem.}
  \textbf{\bibinfo{volume}{68}}, \bibinfo{pages}{549} (\bibinfo{year}{1999}).

\bibitem[{\citenamefont{Parker et~al.}(2002)\citenamefont{Parker, Walker,
  Kendrick, and Pack}}]{Parker-2002}
\bibinfo{author}{\bibfnamefont{G.~A.} \bibnamefont{Parker}},
  \bibinfo{author}{\bibfnamefont{R.~B.} \bibnamefont{Walker}},
  \bibinfo{author}{\bibfnamefont{B.~K.} \bibnamefont{Kendrick}},
  \bibnamefont{and} \bibinfo{author}{\bibfnamefont{R.~T.} \bibnamefont{Pack}},
  \bibinfo{journal}{J. Chem. Phys.} \textbf{\bibinfo{volume}{117}},
  \bibinfo{pages}{6083} (\bibinfo{year}{2002}).

\bibitem[{\citenamefont{Pack et~al.}(1998{\natexlab{b}})\citenamefont{Pack,
  Walker, and Kendrick}}]{Pack-1998b}
\bibinfo{author}{\bibfnamefont{R.~T.} \bibnamefont{Pack}},
  \bibinfo{author}{\bibfnamefont{R.~B.} \bibnamefont{Walker}},
  \bibnamefont{and} \bibinfo{author}{\bibfnamefont{B.~K.}
  \bibnamefont{Kendrick}}, \bibinfo{journal}{J. Chem. Phys.}
  \textbf{\bibinfo{volume}{109}}, \bibinfo{pages}{6714}
  (\bibinfo{year}{1998}{\natexlab{b}}).

\bibitem[{\citenamefont{Smith}(1962)}]{Smith-1962}
\bibinfo{author}{\bibfnamefont{F.~T.} \bibnamefont{Smith}},
  \bibinfo{journal}{Discuss. Faraday Soc.} \textbf{\bibinfo{volume}{33}},
  \bibinfo{pages}{183} (\bibinfo{year}{1962}).

\bibitem[{\citenamefont{Lin}(1995)}]{Lin-1995}
\bibinfo{author}{\bibfnamefont{C.~D.} \bibnamefont{Lin}},
  \bibinfo{journal}{Phys. Pep.} \textbf{\bibinfo{volume}{257}},
  \bibinfo{pages}{1} (\bibinfo{year}{1995}).

\bibitem[{\citenamefont{Esry et~al.}(1996)\citenamefont{Esry, Lin, and
  Greene}}]{Esry-1996}
\bibinfo{author}{\bibfnamefont{B.~D.} \bibnamefont{Esry}},
  \bibinfo{author}{\bibfnamefont{C.~D.} \bibnamefont{Lin}}, \bibnamefont{and}
  \bibinfo{author}{\bibfnamefont{C.~H.} \bibnamefont{Greene}},
  \bibinfo{journal}{Phys. Rev. A} \textbf{\bibinfo{volume}{54}},
  \bibinfo{pages}{394} (\bibinfo{year}{1996}).

\bibitem[{\citenamefont{Aymar et~al.}(1996)\citenamefont{Aymar, Greene, and
  Luc-Koenig}}]{Aymar-1996}
\bibinfo{author}{\bibfnamefont{M.}~\bibnamefont{Aymar}},
  \bibinfo{author}{\bibfnamefont{C.~H.} \bibnamefont{Greene}},
  \bibnamefont{and}
  \bibinfo{author}{\bibfnamefont{E.}~\bibnamefont{Luc-Koenig}},
  \bibinfo{journal}{Rev. Mod. Phys.} \textbf{\bibinfo{volume}{68}},
  \bibinfo{pages}{1015} (\bibinfo{year}{1996}).

\bibitem[{\citenamefont{Tolstikhin et~al.}(1996)\citenamefont{Tolstikhin,
  Watanabe, and Matsuzawa}}]{Tolstikhin-1996}
\bibinfo{author}{\bibfnamefont{O.~I.} \bibnamefont{Tolstikhin}},
  \bibinfo{author}{\bibfnamefont{S.}~\bibnamefont{Watanabe}}, \bibnamefont{and}
  \bibinfo{author}{\bibfnamefont{M.}~\bibnamefont{Matsuzawa}},
  \bibinfo{journal}{J. Phys. B} \textbf{\bibinfo{volume}{29}},
  \bibinfo{pages}{L389} (\bibinfo{year}{1996}).

\bibitem[{\citenamefont{Wang et~al.}(2011)\citenamefont{Wang, D\'{}Incao, and
  Greene}}]{Wang-2011}
\bibinfo{author}{\bibfnamefont{J.}~\bibnamefont{Wang}},
  \bibinfo{author}{\bibfnamefont{J.~P.} \bibnamefont{D\'{}Incao}},
  \bibnamefont{and} \bibinfo{author}{\bibfnamefont{C.~H.}
  \bibnamefont{Greene}}, \bibinfo{journal}{Phys. Rev. A}
  \textbf{\bibinfo{volume}{84}}, \bibinfo{pages}{052721}
  (\bibinfo{year}{2011}).

\bibitem[{\citenamefont{Child}(1974)}]{Child}
\bibinfo{author}{\bibfnamefont{M.~S.} \bibnamefont{Child}},
  \emph{\bibinfo{title}{Molecular Collision Theory}}
  (\bibinfo{publisher}{Academic Press}, \bibinfo{address}{London, England},
  \bibinfo{year}{1974}).

\bibitem[{\citenamefont{Levine}(2005)}]{Levine-2005}
\bibinfo{author}{\bibfnamefont{R.~D.} \bibnamefont{Levine}},
  \emph{\bibinfo{title}{Molecular Reaction Dynamics}}
  (\bibinfo{publisher}{Cambridge University Press},
  \bibinfo{address}{Cambridge, England}, \bibinfo{year}{2005}).

\bibitem[{\citenamefont{Avery}(1989)}]{Avery}
\bibinfo{author}{\bibfnamefont{J.}~\bibnamefont{Avery}},
  \emph{\bibinfo{title}{Hyperspherical Harmonics: Applications in Quantum
  Theory}} (\bibinfo{publisher}{Kluwer}, \bibinfo{address}{Norwell, MA},
  \bibinfo{year}{1989}).

\bibitem[{\citenamefont{Whittaker}(1937)}]{Whittaker-1937}
\bibinfo{author}{\bibfnamefont{E.~T.} \bibnamefont{Whittaker}},
  \emph{\bibinfo{title}{A Trataise on the Analytical Dynamics of Particles and
  Rigid Bodies}} (\bibinfo{publisher}{Cambridge University Press},
  \bibinfo{address}{Cambridge, England}, \bibinfo{year}{1937}).

\bibitem[{\citenamefont{Karplus et~al.}(1965)\citenamefont{Karplus, Porter, and
  Sharma}}]{Karplus-1965}
\bibinfo{author}{\bibfnamefont{M.}~\bibnamefont{Karplus}},
  \bibinfo{author}{\bibfnamefont{R.~N.} \bibnamefont{Porter}},
  \bibnamefont{and} \bibinfo{author}{\bibfnamefont{R.~D.}
  \bibnamefont{Sharma}}, \bibinfo{journal}{J. Chem. Phys.}
  \textbf{\bibinfo{volume}{43}}, \bibinfo{pages}{3259} (\bibinfo{year}{1965}).

\bibitem[{\citenamefont{Aziz et~al.}(1995)\citenamefont{Aziz, Janzen, and
  Moldover}}]{Aziz-1995}
\bibinfo{author}{\bibfnamefont{R.~A.} \bibnamefont{Aziz}},
  \bibinfo{author}{\bibfnamefont{A.~R.} \bibnamefont{Janzen}},
  \bibnamefont{and} \bibinfo{author}{\bibfnamefont{M.~R.}
  \bibnamefont{Moldover}}, \bibinfo{journal}{Phys. Rev. Lett}
  \textbf{\bibinfo{volume}{74}}, \bibinfo{pages}{1586} (\bibinfo{year}{1995}).

\bibitem[{\citenamefont{Press et~al.}(1986)\citenamefont{Press, Teukolsky,
  Vetterling, and Falnnery}}]{Numerical_Recipes}
\bibinfo{author}{\bibfnamefont{W.~H.} \bibnamefont{Press}},
  \bibinfo{author}{\bibfnamefont{S.~A.} \bibnamefont{Teukolsky}},
  \bibinfo{author}{\bibfnamefont{W.~T.} \bibnamefont{Vetterling}},
  \bibnamefont{and} \bibinfo{author}{\bibfnamefont{B.~P.}
  \bibnamefont{Falnnery}}, \emph{\bibinfo{title}{Numerical Recipes in Fotran
  77}} (\bibinfo{publisher}{Cambridge University Press},
  \bibinfo{address}{Cambridge, England}, \bibinfo{year}{1986}).

\bibitem[{\citenamefont{Esry et~al.}(1999)\citenamefont{Esry, Greene, and
  Burke~Jr.}}]{Esry-1999}
\bibinfo{author}{\bibfnamefont{B.~D.} \bibnamefont{Esry}},
  \bibinfo{author}{\bibfnamefont{C.~H.} \bibnamefont{Greene}},
  \bibnamefont{and} \bibinfo{author}{\bibfnamefont{J.~P.}
  \bibnamefont{Burke~Jr.}}, \bibinfo{journal}{Phys. Rev. Lett}
  \textbf{\bibinfo{volume}{83}}, \bibinfo{pages}{1751} (\bibinfo{year}{1999}).

\bibitem[{\citenamefont{Suno et~al.}(2002)\citenamefont{Suno, Esry, Grene, and
  Burke~Jr.}}]{Suno-2002}
\bibinfo{author}{\bibfnamefont{H.}~\bibnamefont{Suno}},
  \bibinfo{author}{\bibfnamefont{B.~D.} \bibnamefont{Esry}},
  \bibinfo{author}{\bibfnamefont{C.~H.} \bibnamefont{Grene}}, \bibnamefont{and}
  \bibinfo{author}{\bibfnamefont{J.~P.} \bibnamefont{Burke~Jr.}},
  \bibinfo{journal}{Phys. Rev. A} \textbf{\bibinfo{volume}{65}},
  \bibinfo{pages}{042725} (\bibinfo{year}{2002}).

\bibitem[{\citenamefont{Wang et~al.}(2012)\citenamefont{Wang, D\'{}Incao, Wang,
  and Greene}}]{Wang-2012}
\bibinfo{author}{\bibfnamefont{J.}~\bibnamefont{Wang}},
  \bibinfo{author}{\bibfnamefont{J.~P.} \bibnamefont{D\'{}Incao}},
  \bibinfo{author}{\bibfnamefont{Y.}~\bibnamefont{Wang}}, \bibnamefont{and}
  \bibinfo{author}{\bibfnamefont{C.~H.} \bibnamefont{Greene}},
  \bibinfo{journal}{Phys. Rev. A} \textbf{\bibinfo{volume}{86}},
  \bibinfo{pages}{062511} (\bibinfo{year}{2012}).

\bibitem[{\citenamefont{Yu.~Schreiber}(1966)}]{Schreiber-1966}
\bibinfo{editor}{\bibfnamefont{A.}~\bibnamefont{Yu.~Schreiber}}, ed.,
  \emph{\bibinfo{title}{The Monte Carlo Method: The Method for Statistical
  Trials}} (\bibinfo{publisher}{Pergamon Press}, \bibinfo{address}{Oxford},
  \bibinfo{year}{1966}).

\bibitem[{\citenamefont{Landau and Binder}(2009)}]{Landau-2009}
\bibinfo{author}{\bibfnamefont{D.~P.} \bibnamefont{Landau}} \bibnamefont{and}
  \bibinfo{author}{\bibfnamefont{K.}~\bibnamefont{Binder}},
  \emph{\bibinfo{title}{A guide to Monte-Carlo Simulations in Statistical
  Physics}} (\bibinfo{publisher}{Cambridge University Press},
  \bibinfo{address}{Cambridge, England}, \bibinfo{year}{2009}).

\bibitem[{\citenamefont{Truhlar and Muckerman}(1975)}]{Truhlar-1975}
\bibinfo{author}{\bibfnamefont{D.~G.} \bibnamefont{Truhlar}} \bibnamefont{and}
  \bibinfo{author}{\bibfnamefont{J.~T.} \bibnamefont{Muckerman}}, in
  \emph{\bibinfo{booktitle}{Atom-Molecule Collision Theory: A Guide for the
  Experimentalist}}, edited by \bibinfo{editor}{\bibfnamefont{R.~B.}
  \bibnamefont{Bernstein}} (\bibinfo{publisher}{Plenum Press, New York},
  \bibinfo{year}{1975}), chap.~\bibinfo{chapter}{16}, pp.
  \bibinfo{pages}{505--566}.

\bibitem[{\citenamefont{Shui}(1972)}]{Shui-thesis}
\bibinfo{author}{\bibfnamefont{V.~H.} \bibnamefont{Shui}}, Ph.D. thesis,
  \bibinfo{school}{Deparment of mechanical ingeneering MIT}
  (\bibinfo{year}{1972}).

\bibitem[{\citenamefont{Shui}(1973)}]{Shui-1973}
\bibinfo{author}{\bibfnamefont{V.~H.} \bibnamefont{Shui}}, \bibinfo{journal}{J.
  Chem. Phys.} \textbf{\bibinfo{volume}{58}}, \bibinfo{pages}{4868}
  (\bibinfo{year}{1973}).

\bibitem[{\citenamefont{Landau and Lifshitz}(1958)}]{Landau}
\bibinfo{author}{\bibfnamefont{L.~D.} \bibnamefont{Landau}} \bibnamefont{and}
  \bibinfo{author}{\bibfnamefont{E.~M.} \bibnamefont{Lifshitz}},
  \emph{\bibinfo{title}{Quantum Mechanics}} (\bibinfo{publisher}{Butterworht
  Heinemann}, \bibinfo{address}{Oxfort, England}, \bibinfo{year}{1958}).

\bibitem[{\citenamefont{Kendrick et~al.}(1999)\citenamefont{Kendrick, Pack,
  Walker, and Hayes}}]{Kendrick-1999}
\bibinfo{author}{\bibfnamefont{B.~K.} \bibnamefont{Kendrick}},
  \bibinfo{author}{\bibfnamefont{R.~T.} \bibnamefont{Pack}},
  \bibinfo{author}{\bibfnamefont{R.~B.} \bibnamefont{Walker}},
  \bibnamefont{and} \bibinfo{author}{\bibfnamefont{E.~F.} \bibnamefont{Hayes}},
  \bibinfo{journal}{J. Chem. Phys.} \textbf{\bibinfo{volume}{110}},
  \bibinfo{pages}{6673} (\bibinfo{year}{1999}).

\bibitem[{\citenamefont{Lepetit et~al.}(1990)\citenamefont{Lepetit, Peng, and
  Kuppermann}}]{Lepetit-1990}
\bibinfo{author}{\bibfnamefont{B.}~\bibnamefont{Lepetit}},
  \bibinfo{author}{\bibfnamefont{Z.}~\bibnamefont{Peng}}, \bibnamefont{and}
  \bibinfo{author}{\bibfnamefont{A.}~\bibnamefont{Kuppermann}},
  \bibinfo{journal}{Chem. Phys. Lett.} \textbf{\bibinfo{volume}{166}},
  \bibinfo{pages}{572} (\bibinfo{year}{1990}).

\bibitem[{\citenamefont{Suno and Esry}(2008)}]{Suno-2008}
\bibinfo{author}{\bibfnamefont{H.}~\bibnamefont{Suno}} \bibnamefont{and}
  \bibinfo{author}{\bibfnamefont{B.~D.} \bibnamefont{Esry}},
  \bibinfo{journal}{Phys. Rev. A} \textbf{\bibinfo{volume}{78}},
  \bibinfo{pages}{062701} (\bibinfo{year}{2008}).

\bibitem[{\citenamefont{Mott and Massey}(1965)}]{Mott}
\bibinfo{author}{\bibfnamefont{N.~F.} \bibnamefont{Mott}} \bibnamefont{and}
  \bibinfo{author}{\bibfnamefont{H.~S.~W.} \bibnamefont{Massey}},
  \emph{\bibinfo{title}{The Theory of Atomic Collisions}}
  (\bibinfo{publisher}{Claredon Press}, \bibinfo{address}{Oxford},
  \bibinfo{year}{1965}), \bibinfo{edition}{3rd} ed.

\bibitem[{\citenamefont{Mehta et~al.}(2009)\citenamefont{Mehta, Rittenhouse,
  D\'{}Incao, von Stecher, and Greene}}]{Mehta-2009}
\bibinfo{author}{\bibfnamefont{N.~P.} \bibnamefont{Mehta}},
  \bibinfo{author}{\bibfnamefont{S.~T.} \bibnamefont{Rittenhouse}},
  \bibinfo{author}{\bibfnamefont{J.~P.} \bibnamefont{D\'{}Incao}},
  \bibinfo{author}{\bibfnamefont{J.}~\bibnamefont{von Stecher}},
  \bibnamefont{and} \bibinfo{author}{\bibfnamefont{C.~H.}
  \bibnamefont{Greene}}, \bibinfo{journal}{Phys. Rev. Lett}
  \textbf{\bibinfo{volume}{103}}, \bibinfo{pages}{153201}
  (\bibinfo{year}{2009}).

\bibitem[{\citenamefont{Bruch et~al.}(2002)\citenamefont{Bruch, Sch\"{o}llkopf,
  and Toennies}}]{Bruch-2002}
\bibinfo{author}{\bibfnamefont{L.~W.} \bibnamefont{Bruch}},
  \bibinfo{author}{\bibfnamefont{W.}~\bibnamefont{Sch\"{o}llkopf}},
  \bibnamefont{and} \bibinfo{author}{\bibfnamefont{J.~P.}
  \bibnamefont{Toennies}}, \bibinfo{journal}{J. Chem. Phys.}
  \textbf{\bibinfo{volume}{117}}, \bibinfo{pages}{1544} (\bibinfo{year}{2002}).

\bibitem[{\citenamefont{Jones et~al.}(2006)\citenamefont{Jones, Tiesinga, Lett,
  and Julienne}}]{Jones-2006}
\bibinfo{author}{\bibfnamefont{K.~M.} \bibnamefont{Jones}},
  \bibinfo{author}{\bibfnamefont{E.}~\bibnamefont{Tiesinga}},
  \bibinfo{author}{\bibfnamefont{P.~D.} \bibnamefont{Lett}}, \bibnamefont{and}
  \bibinfo{author}{\bibfnamefont{P.~S.} \bibnamefont{Julienne}},
  \bibinfo{journal}{Rev. Mod. Phys.} \textbf{\bibinfo{volume}{78}},
  \bibinfo{pages}{483} (\bibinfo{year}{2006}).

\bibitem[{\citenamefont{Shepard}(2007)}]{Shepard-2007}
\bibinfo{author}{\bibfnamefont{J.~R.} \bibnamefont{Shepard}},
  \bibinfo{journal}{Phys. Rev. A} \textbf{\bibinfo{volume}{75}},
  \bibinfo{pages}{062713} (\bibinfo{year}{2007}).

\bibitem[{\citenamefont{Rittenhouse et~al.}(2011)\citenamefont{Rittenhouse, von
  Stecher, D\'{}Incao, Mehta, and Greene}}]{Rittenhouse-2011JPB}
\bibinfo{author}{\bibfnamefont{S.~T.} \bibnamefont{Rittenhouse}},
  \bibinfo{author}{\bibfnamefont{J.}~\bibnamefont{von Stecher}},
  \bibinfo{author}{\bibfnamefont{J.~P.} \bibnamefont{D\'{}Incao}},
  \bibinfo{author}{\bibfnamefont{N.~P.} \bibnamefont{Mehta}}, \bibnamefont{and}
  \bibinfo{author}{\bibfnamefont{C.~H.} \bibnamefont{Greene}},
  \bibinfo{journal}{J. Phys. B} \textbf{\bibinfo{volume}{44}},
  \bibinfo{pages}{172001} (\bibinfo{year}{2011}).

\bibitem[{\citenamefont{Wang et~al.}(2013)\citenamefont{Wang, D\'{}Incao, and
  Esry}}]{DIncao-2013AdvAtMolecPhys}
\bibinfo{author}{\bibfnamefont{Y.}~\bibnamefont{Wang}},
  \bibinfo{author}{\bibfnamefont{J.~P.} \bibnamefont{D\'{}Incao}},
  \bibnamefont{and} \bibinfo{author}{\bibfnamefont{B.~D.} \bibnamefont{Esry}},
  \bibinfo{journal}{Adv. At. Mol. Phys.} \textbf{\bibinfo{volume}{62}},
  \bibinfo{pages}{1} (\bibinfo{year}{2013}).

\end{thebibliography}

\end{document}